\documentclass{aastex}          
\usepackage{spr-astr-addons}    

\begin{document}
%
\title{Habitability on local, Galactic and cosmological scales}

\shorttitle{Habitability}
\shortauthors{<Fecchio et al.>}

\author{Luigi Secco\altaffilmark{1}} 
\author{Marco Fecchio\altaffilmark{1}}
\email{luigi.secco@unipd.it} 
\author{Francesco Marzari\altaffilmark{1}}
\altaffiltext{1}{Department of Physics \& Astronomy, Padova University}

\begin{abstract}
The aim of this paper is to underline conditions necessary for the emergence and development of life. They are placed at local planetary scale, at Galactic scale and within the cosmological evolution, as pointed out by the Anthropic Cosmological Principle. We will consider the circumstellar habitable zone (CHZ) for planetary systems and a Galactic Habitable Zone (GHZ) including also a set of strong cosmological constraints to allow life (cosmological habitability (COSH)). 
Some requirements are specific of a single scale and its related physical phenomena, while others are due to the conspired effects occurring  at more than one scale. The scenario emerging from this analysis is that all the habitability conditions here detailed must at least be met. Then life asks something which may appear as "a monstrous sequence of accidents" as Hoyle (1959) thought, or as providential collaborations leading to realize
how finely tuned is the architecture within which Life, as precious stone, is embedded.
\end{abstract}


%
\section{Introduction}
The first step is to specify to what life we refer. This leads to the essential requests that must be fulfilled to allow the development of  this kind of life (sect.2) and to which astrophysical scales they mainly are related (sect.3).
In sect.4 we point out how in the Solar system, among ten planets and hundreds of satellites, life in complex form appears only on Earth, proving to be peculiar without being probably exclusive. Some of the many constraints on arising and developing of it may be detectable studying our site. The Climatic Astronomical Theory is introduced in sect.5 in order to define the circumsolar habitable zone (HZ) (sect.6) while the translation from Solar to extra-Solar systems leads to a generalized circumstellar habitable zone (CHZ) defined in sect.7 with some exemplifications to the Gliese-667C and the TRAPPIST-1 systems; some general remarks follow (sect.8). A first conclusion related to CHZ is done moving toward GHZ and COSH (sect.9). The conditions for the development of life are indeed only partially connected to the local scale in which a planet is located. A strong interplay between different scales exists and each single contribution to life from individual scales is difficult to be isolated. However we will limit ourselves to a first order approach in which the analyzed necessary ingredients for life are first of all linked to a given main scale, taking the more complex connections with the others as considerations of second order.\\ 
In sect.10 we will introduce the Galactic Habitable Zone (GHZ) to which mostly leads the following requests: i) the metal amount necessary to form a planet like Earth, ii) quantify the threats of supernovae explosions. The main contributions from different authors on the GHZ limits are also analyzed. Then we consider the role of possible comet injections from Oort's Cloud (or analogs) due to tidal effects coming from the main dynamical Galaxy components (sect.11). In sect.12 we add a new constraint to GHZ due to the possibility of a comet flux with relevant consequence for life. That is bound to the location of planetary system on Galactic plane. Further a probable Solar path through the Galaxy has to be considered owing to the Sun's over-metallicity. In this way we will be forced to move to the wider history of Galaxy formation and to its chemical-dynamical evolution. Some paths have been simulated (Kaib et al., 2011) and overlapped to GHZ (sect.13). Then the horizon gets larger to the cosmology where many fine tunings exist in the connection between some factors constraining the main features and evolution of Universe and life. The set of these fine tuned cosmological constraints define, in our view, the cosmological habitability (COSH) (sect.14). 
Conclusions are given in sect.15. 
 
\section{What kind of life?}
What kind of life are we referring to? We focus on the typical life on Earth, the only one we know so far, therefore we consider environments suitable to aerobic complex life based on oxigen and carbon excluding the wide spectrum of life-forms limited at micro-organism level.
In addition, all organisms we are familiar with require liquid water during at least part of their life cycle.
Here below we summarize the essential requests for this kind of life in order of decreasing priority \citep{das93} and putting in brackets the scales to which they mainly refer (see, sect.3):
\begin{itemize}
\item $\alpha )$ the presence of bricks, it means the chemical elements which are basilar in order to form organic compounds. As mentioned in Barrow and Tipler (1986), the biochemist Lehninger (1975) divided the chemical elements important for life into three main classes. In class one there are the essential elements which are present (for one per cent or more) in all living organisms (with the exception of sulphur). They are oxygen, carbon, nitrogen, hydrogen, phosphorus, and sulphur. In class two it appears the monoatomic ions, such as $N_a^+$ and $K^+$, in class three the trace elements: iron, copper and so one.\\
In COSH, we will take into account how special is the tuning of coupling constants in producing primordial helium at the beginning of Universe and how carbon and oxygen burnings occur later inside the first star generation. In addition, in Appendix we will outline the main nuclear reactions able to produce nitrogen and phosphorous inside stars (COSH).
\begin{figure}[tb]
\includegraphics[width=0.95\columnwidth]{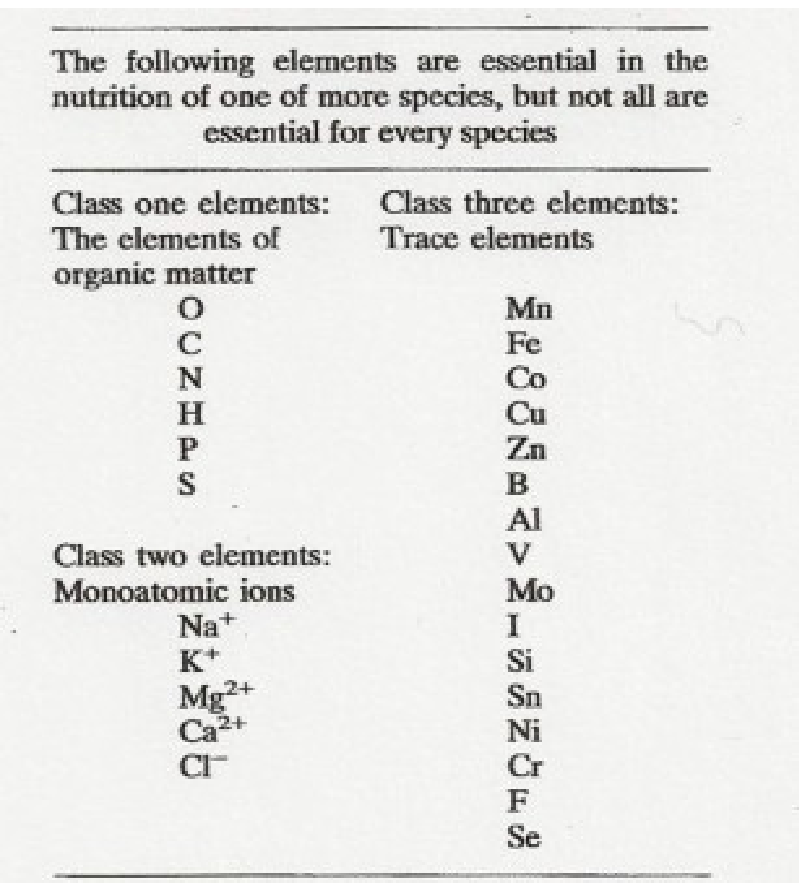}
\caption{The elements used in living organisms, see text, Barrow and Tipler (1986), pg.553.}
\end{figure}

 \item $\beta )$ A sufficiently long time as the terrestrial experience teaches us. Indeed the earliest micro-fossil evidence for life is about 3.5 billion years old, that is at least $\simeq 1\cdot 10^9 yr$ after the formation of the planet. They are unicellular and prokaryote organisms from some of the oldest rocks on the planet, similar to bacteria and to blue green algae. Moreover to reach complex, pluricellular eukaryotic forms of life it is necessary to add another $\simeq 2.7\cdot 10^9 yr$ (Fig.1)(Simpson, 1986, pg.71; Curtis \& Barnes, 1991; for an updated and detailed timetable of life developing in: Eons, Eras, Periods and Ages, see, Knoll \& Nowak, 2017) (GHZ, COSH).
\item $\gamma )$ Planetary conditions suitable to harbour life. This means orbital stability leading to steady atmospheric temperature and pressure (organic compounds have to find not only the possibility to form but also to be stable). Their values must also be compatible with the presence of liquid water on the surface (CHZ).
\item $\delta )$ In addition, the environment has  to be suitable to form long molecular biological chains (CHZ and GHZ).
\end{itemize}  
Human Life might request more stringent constraints. 

\begin{figure}[tb]
\includegraphics[width=1.0\columnwidth]{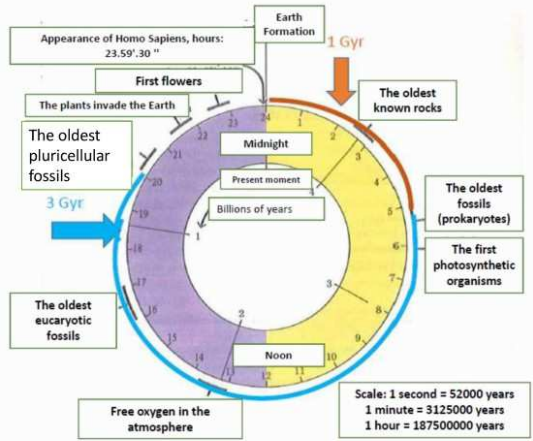}
\caption{The biological time clock: the age of Solar system (4.5 Gyr) is divided into 24 hours, 60 minutes and seconds. The correspondence is shown in the legend. After Earth formation primordial unicellular and prokaryote organisms needed  at least $\simeq 1 \cdot 10^9 yr$ to appear (before six a.m.; red line and arrow). To reach complex pluricellular eukaryotic forms of life we need to add again about 3 Gyr (around 8 p.m.; blue line and arrow)
 (adapted from:\textit{Curtis \& Barnes, 1991).}}
\label{OROLOGIO} 
\end{figure}

\section{The different scales}
Each of the essential requests: $\alpha), \beta ), \gamma ), \delta )$ are mainly linked to the 
physical phenomena occurring at one of the three scales of habitability considered:\\
\begin{itemize}
\item  Local or planetary: it defines a circumstellar habitable zone (CHZ) on the basis of the \textit{greenhouse effect} and the presence of liquid water.
\item  Galactic: it deals with the metal amount needed to obtain a terrestrial planet, the effects of supernovae explosions, the Galactic tide on comets and the necessary time for developing complex life. These constraints leads to the definition of Galactic habitable zone (GHZ).
\item  Cosmological: it involves the relationship between life and cosmological evolution and it outlines the cosmological habitability (COSH). 
\end{itemize}
It frequently happens that constraints from one scale correlate to those of another one making it difficult to disintangle the individual contribution. In sect.2, we have put in brackets some indications to drive the reader through the items. 

\begin{figure}[!]
\includegraphics[width=1.0\columnwidth,angle=0.0]{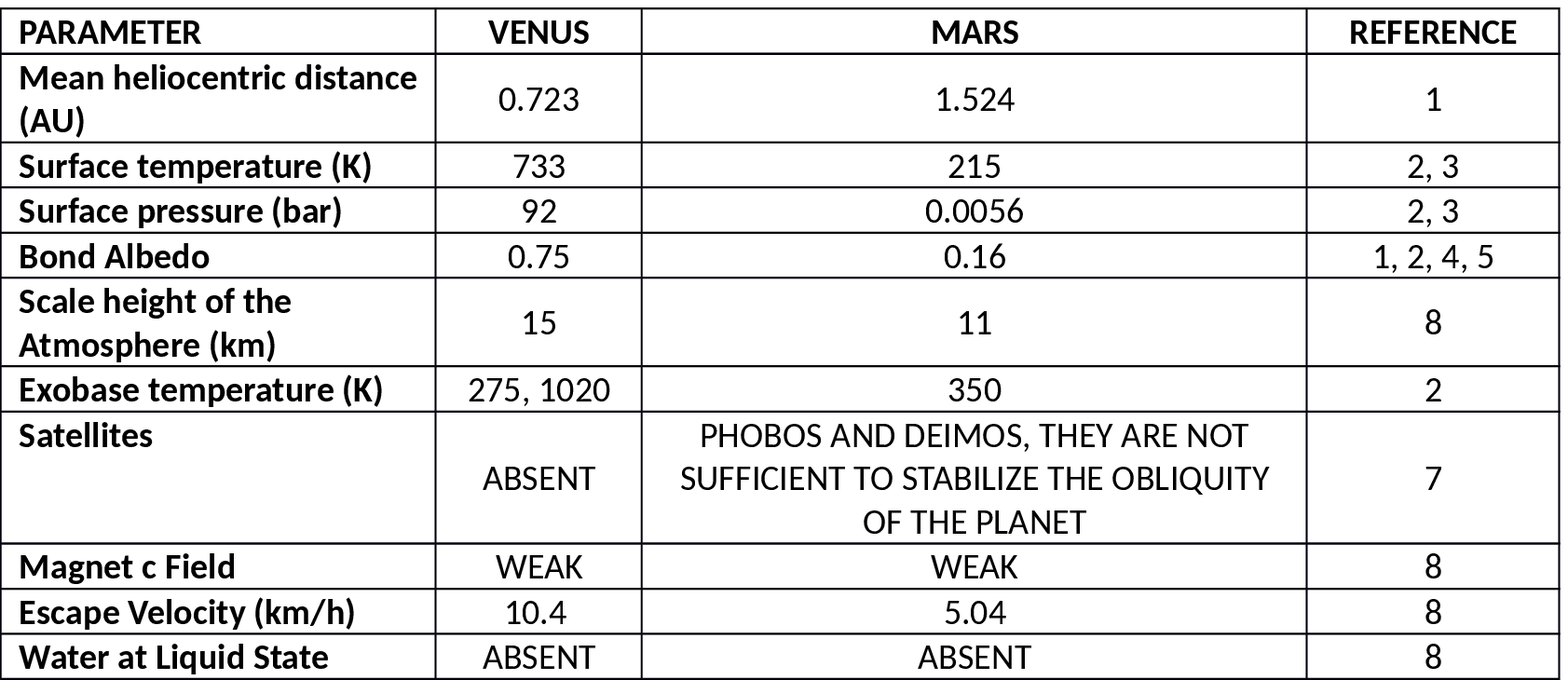}
\caption{Collections of different characteristics related to our neighboors: Mars and Venus (references: 1, Yoder (1995); 2, Chamberlain \& Hunten (1987); 3, Clarke et al. (1992); 4, Veverka et al. (1988); 5, Moroz (1983); 6, Stern and Yelle (1999), in De Pater e Lissauer (2001);
7, Tomasella et al. (1996); 8, Marzari \& Vanzani (2013)). 
 }
\label{COLL}
\end{figure}

\section{The Earth peculiarity}
 We collect here some features that allowed the Earth to reach suitable conditions for life and compare them with those of our neighbooring planets (Fig.\ref{COLL}).\\
The aim doesn't go in the direction to exclude other possible life like that on the Earth, but realistically to analyze the necessary constraints to allow it. Peculiarity indeed doesn't mean exclusivity. According to the Weak Anthropic Principle (see, sect.15) the fine tuned conditions of the Universe are in fact to obtain life, as the only one we know, but spread in the whole Universe. 
\begin{itemize}
\item Its revolution around the Sun is placed at a distance (1 AU) that permits the presence of water in a liquid, solid and gaseous state on its surface;
\item its atmosphere is mainly composed by nitrogen and oxygen;
\item the intensity of its magnetic field (0.3-0.4 G) is the highest within the Terrestrial Planets so that it can act as a shield against cosmic rays and solar wind;
\item among the Terrestrial Planets, Earth is the most massive and the escape velocity turns out to be very high (11.2 km/sec). The consequence is a stable atmosphere with a reduced thermal loss of atmospheric components.
\item Its only satellite, the Moon, is massive ($\simeq 1/80 \ M_{\oplus}$ ; the highest ratio in the Ss) and close enough ($\simeq 30 \ d_{\oplus}$) to stabilize the Earth obliquity allowing a regular succession of seasons on a very long time-scale (see, sect.5.2 for a wider discussion).
\item The existence of Earth's plate tectonics which has a crucial role in the geochemical carbonate - silicate cycle as discussed in the next sect. 6.1.1. From the preliminary results of Noack \& Breuer (2011) it seems that propensity of plate tectonics may occur more seldom than previously thought with a peak between one and five Earth masses. For Mars, in comparison, there are a number of lines of circumstantial evidence to suggest that it did in the past have some form of surface tectonics or tectonic episodes but the problem remains completely unresolved (Zhang \& O'Neil, 2016). Moreover, about the volcanism which is essential part of geochemical cycle, thermal evolution models of Mars suggest that its volcanism should have stopped at least some hundred million years ago (Hauck and Phillips, 2002; Breuer and Spohn, 2003), although it is possible that Mars is not yet volcanically extinct as claimed by Hauber et al., 2011. 
\end{itemize}

\begin{figure}
\begin{center}
\includegraphics[width=5.0cm]{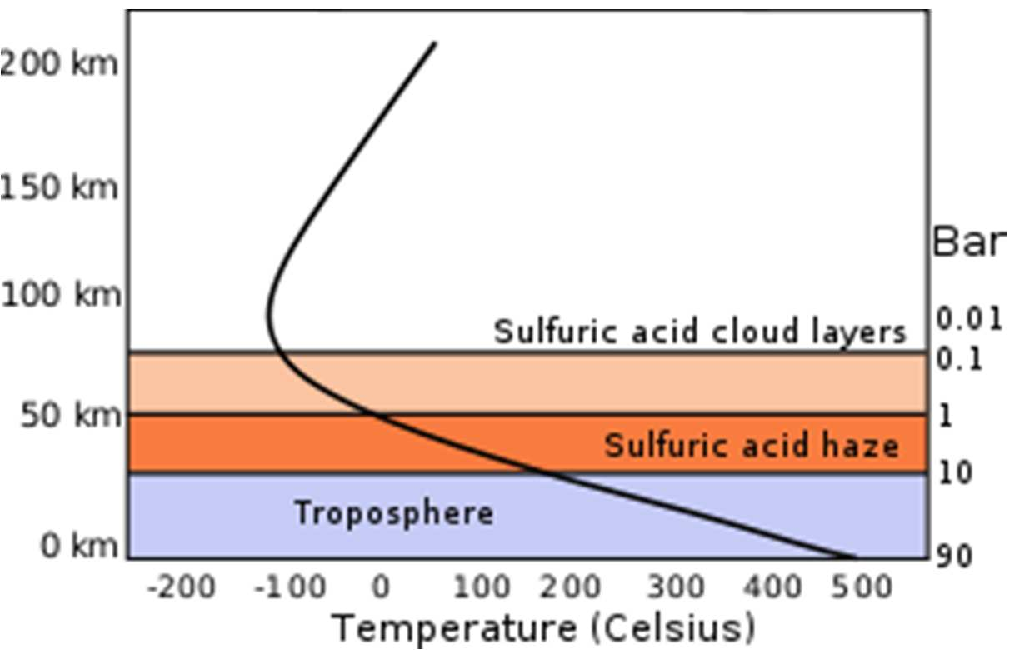}
\includegraphics[width=7.0cm]{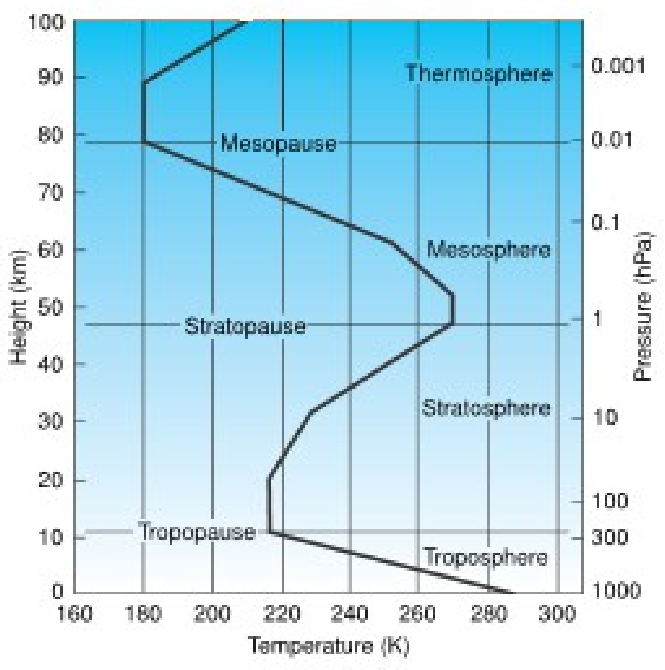}
\includegraphics[width=7.0cm]{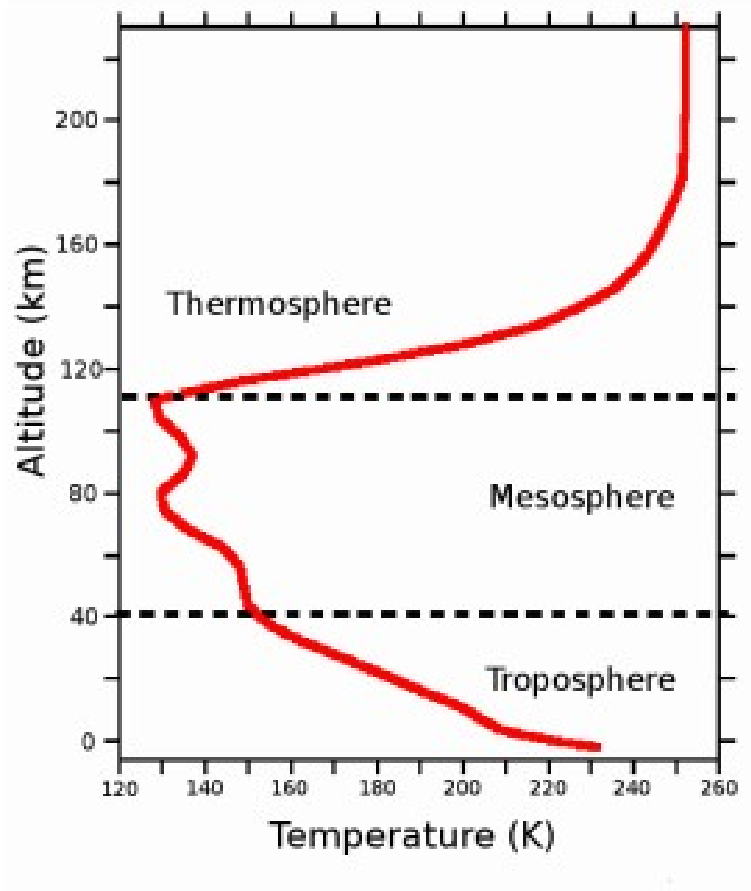}
\end{center}
\caption{From top to bottom: Venus, Earth and Mars' temperature gradients. Credits: Palen et al. (2011); Wallace \& Hobbs, (2006) ; Gonz\'alez-Galindo et al. (2008), respectively.}   
\label{VENUS}
\label{EARTH}
\label{MARS}
\end{figure}



In Fig.4 Earth, Mars and Venus' temperature gradients are also compared. They become relevant in determining the greehouse effect (e.g. the tendency of water vapor to escape from the Earth is minimal while this is not the case for the early Venus (Kasting et al., 1988)). 

\begin{figure}[!]
\includegraphics[width=0.7\columnwidth,angle=0]{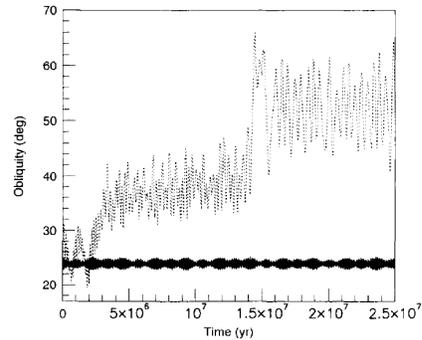}
\caption{Evolution of Earth's obliquity, $\theta$, for the real Earth-Moon system (continuous line) and in a test case where Moon's distance from Earh is increased from 60.3 to about 66.5 terrestrial radii (dotted line) (Tomasella et al., 1996).}
\label{FIG5}
\end{figure}

\begin{figure}[!]
\includegraphics[width=0.8\columnwidth,angle=0]{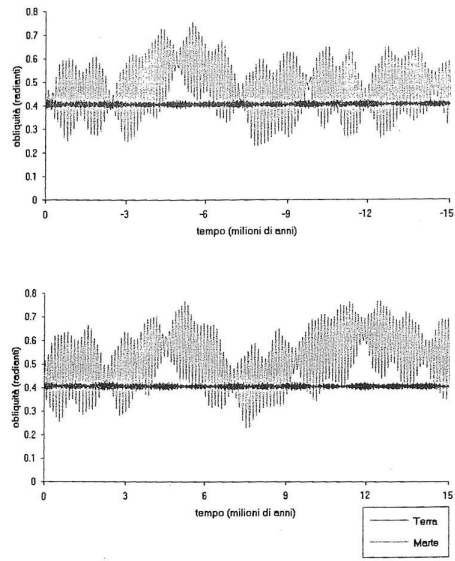}
\caption{Comparison between the variations of obliquity $\theta$ of Mars in the Mars-Phobos-Deimos system (dotted line) vs that of Earth in the Earth-Moon system (continuous line) for 15 million years before (upper) and after present (down) (simulations by Tomasella, 1992). Mars maximum oscillations are of $\pm 15.4^o$ around the mean value of $23.4^o$.}
\label{FIG6}
\end{figure}

\section{Relationship between climatic history and dynamics}
Since the beginning of 1900, various models relating the climatic history of a planet with its dynamical evolution within Ss have been formulated leading to the so called Climatic Astronomical Theory (e.g., see, Ward, 1974).\\ 
Ad example, Earth's mean temperature depends not only on the Earth-Sun distance which determines the Solar flux on the planet, but it is also subordinate (as we will see) to its dynamics and to the climate model for its atmosphere (sect.6) (Kasting et al., 1988; Kasting et al., 1993).
 
\subsection{Orbital elements}
Usually, the main orbital elements relevant in the determination of the planet's temperature 
are the eccentricity $e$, longitude of perihelion $\varpi$ and the semimajor axis $a$. In addition, also the obliquity $\theta$ has a relevant role. The average \textit{insolation} during a periodic revolution, can be estimated at the planet's pole (Ward, 1974) as:
\begin{equation}
\label{insol}
\left\langle I \right\rangle = S \ sin \theta/ \pi \ (1 - e^2)^{-1/2}
\end{equation}
where $S$ is the Solar constant, $S_o$, scaled at the planet's distance in AU.\\

\subsection{On Earth's obliquity}
Due to the fast rotation around its axis, the Earth shape becomes spheroidal with an equatorial semimajor axis larger than the semiminor one of about 21 km. This causes a precession of its rotational axis due to the Luni-Solar effect with a period equal to about $T_{p\oplus}\simeq 26,000 \ yrs$. It is the fastest precession rate in the Ss (e.g., for Mars, $T_{pM}\simeq 175,000 \ yrs$),
placing at present the Earth far from 
resonances between its precession motion 
and any fundamental frequency of the secular motion 
of the Ss which might cause  a chaotic
evolution of the obliquity of the planet (Laskar \& Robutel, 
1993; Tomasella et al., 1996; Neron de Surgy \&
Laskar, 1997).
The occurrence of 
these resonances depend on the architecture of the planetary 
system which sets the fundamental frequencies of its secular
evolution and the precession rate of the planet spin axis which 
may be determined by the presence of a satellite. In the case of the 
Earth, the  Moon
may lead in the future the Earth into a chaotic state when the tidal 
interaction between the two bodies will have moved the Moon
to 66.5 Earth's radii (it is currently at 60).
This is shown in Fig.\ref{FIG5} illustrating the evolution of the Earth obliquity as the Moon moves from its current distance of 60.3 terrestrial radii (continuous line) to 66.5 (dotted line) (Tomasella et al., 1996). This behavior is compared to that of Mars (Fig.\ref{FIG6})  where its obliquity cannot avoid, due to the small masses of Phobos and Deimos, variations of its rotazional axis up to $\pm 15.4^o$ around a mean value of $23.4^o$.
This behavior leads to an important 
requirement for a planet to be habitable i.e. that its precession
rate be far from any fundamental frequency of the system 
whose values are determined by the mass and semimajor axes 
of all planets in the system. \\

\subsection{The greenhouse effect}
\begin{figure}[!]
\includegraphics[width=1.0\columnwidth,angle=00]{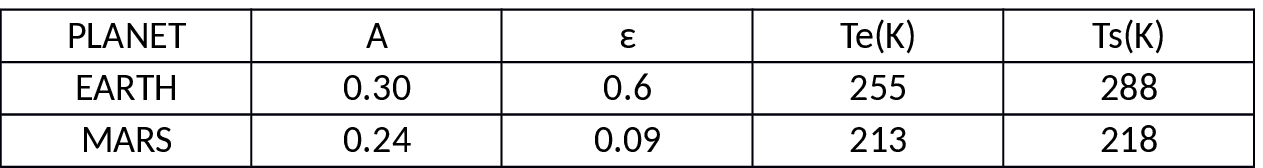}
\caption{Comparison of global \textit{albedo} A (in visible range), emissivity $\epsilon$, effective temperatures $T_e$ and surface temperatures $T_s$, between Mars and Earth (Eqs. \ref{TB}, \ref{TBS}, \ref{Ort12}) for a simplified atmospheric model (see, text) (Ortolani, 2012; Pall\'e et al. 2016).}
\label{TETS}
\end{figure}

The temperature of a planet surface is strongly depending on \textit{the greenhouse effect} due to the planet atmosphere.
To roughly quantify its relevance, let us remember the approximate results for Earth and Mars, collected in Fig.\ref{TETS}, obtained by a simplified atmospheric model (Ortolani, 2012; Davolio, 2012).
The solar flux incident at the top of the atmosphere is equal to the present solar constant at Earth's orbit, $S_o= 1360 \ W \ m^{-2}$. The radiative balance between the flux coming in from Sun, $S$, and that emitted from planet surface of radius $R$, as a black-body with effective temperature $T_e$, without atmosphere, reads as:
 
\begin{equation}
\label{TB}
S=(1-A)\frac{S_o}{4}= \sigma T_e^4
\end{equation}
$\sigma$ being the Stefan-Boltzmann constant, $A$ the planetary (Bond) albedo. If $T_s$ is the surface temperature when atmosphere is present and $T_A$ the temperature of a single \textit{greenhouse gas layer}, the inclusion of its contribution, transforms (\ref{TB}) into:

\begin{equation}
\label{TBS}
(1-A) \frac{S_o}{4}+ \epsilon \sigma T_A^4=\sigma T_s^4\\
\end{equation}
because the atmosphere radiates equally outwards and inwards.
The relationship between $T_e$ and $T_s$ becomes:
\begin{equation}
\label{Ort12}
\frac{T_s^4}{T_e^4}=\frac{1}{1-\epsilon/2}
\end{equation}
Where $\epsilon$ is the \textit{emissivity} of the atmosphere.

\subsubsection {To quantify greenhouse effect}
The present Earth's atmosphere is done, as percentage by weight, of nitrogen, $78 \%$, of oxygen, $21 \%$, of argon $1 \%$, of carbon dioxide, $0.03 \%$.\\
The fraction of   
Solar flux reflected by the high part of the atmosphere, for Earth (in the visible range) is: $A \sim 0.3$ (Pall\'e et al. 2016). 
Many factors are involved in determining both \textit{albedo} and the emissivity $\epsilon$ of an atmosphere:
\begin{itemize}
\item the amount of lands;
\item the amount of condensed volatile substances (extension of ice caps, of liquid fraction and cloud conformation).
\end{itemize}
 
For the Earth, the UV radiation, in the range $200 \div 300 nm$, is absorbed by the ozone in the \textit{stratosphere} while the IR component is absorbed by the \textit{troposphere} which in turn diffuses it inside thanks to the presence of water vapor and $CO_2$. The amounts of both these components determine the \textit{greenhouse effect} (Fig.\ref{QUAL}).\\


\section{Circumstellar Habitable Zone for Solar system (HZ)}
\subsection{Standard climate Model}
In order to estimate the width of the HZ around stars similar to our own Sun we introduce the standard climate model for the Earth and Terrestrial Planets in general, as proposed by Kasting et al. (1988), Kasting et al.(1993). In this model the atmospheres are made of $CO_2/H_2O/N_2$.\\
The vapor pressure of water at the surface is a function of surface temperature. In most of the calculations of this 1-D model, $N_2$ is at 1 bar of pressure while $CO_2$ has a variable partial pressure and $H_2O$ and $CO_2$ clouds are not included explicitely.\\
\subsubsection{The role of carbon dioxide}
Due to its four electrons in the L shell carbon form a double bond with oxigen in $CO_2$. Since each oxygen atom has a valence of two, the double bond completes the L shell of each oxygen atom, and carbon dioxide molecule has very little remaining affinity for itself. This means that carbon dioxide is a gas at ordinary temperatures.\footnote{That doesn't occur,e.g., for silicon dioxide which is a solid, even though silicon is very similar to carbon and has the same valence of four.}\\
$CO_2$ and its evolution is governed by carbonate-silicate geochemical cycle. Following Kasting et al. (1988) and Kasting (2001), we may summarize it as follows: $CO_2$ is emitted from volcanoes and from atmosphere it dissolves in rainwater which erodes rocks that contain calcium-silicate minerals ($C_aSiO_3$). In the chemical process calcium and bicarbonate ions ($Ca^{++}$ and $HCO_3^-$) are released into the groundwater and ultimately transported to the ocean.
In the sea, plankton and other organisms incorporate them. When the organisms die, they will form the carbonate sediments at the bottom of the ocean. When seafloor is subducted as a result of plate tectonics activity, calcium carbonate ($C_aCO_3$) subjected to rising temperature and pressure reacts with silica re-forming silicate rocks and releasing gaseous carbon dioxide by , e.g., volcanic eruptions. As highlighted by Kasting et al. (1988) that is a feedback system owing to the Earth's temperature has always remained within reasonable limits even though for stellar evolution Sun changed its luminosity of about $(25 \div 30)\%$ during the $4.5$ billion years from the formation of solar system. \textit{Thus the carbonate-silicate cycle acts as a planetary thermostat that regulates mean surface temperatures} (Ramirez, 2018). Without it the Earth's atmosphere would be only marginally stable, which means it could be destabilized by relatively small perturbations. That indeed occurs in the Hart's computer simulations (Hart, 1978.) According to them, as the free oxygen concentration gradually rose from zero in the beginning to the current level due to phothosynthesis of green plants the greenhouse effect faded away with the effect that the temperature decreased drastically (Barrow \& Tipler, 1986, Chapt.8).\\    
\subsubsection{The effective Solar flux} 
\begin{figure}[tb]
 \includegraphics[width=0.90\columnwidth,angle=00]{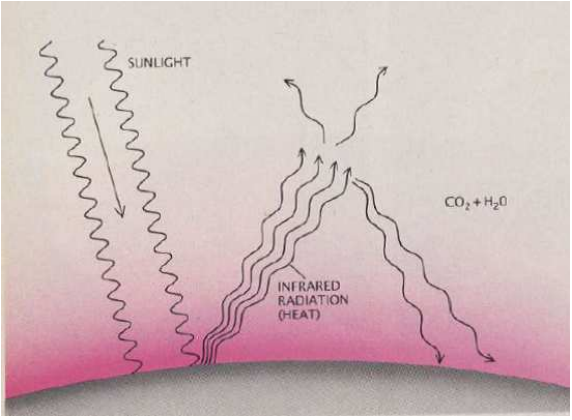}
\caption{Qualitative picture to understand the energy balance (\ref{SE}) when the atmosphere produces  \textit{greenhouse effect}. Carbon dioxide and water vapor intercept the infrared rays (heat) that planet radiates into space and reradiate much of this energy toward the surface. Only part of this infrared radiation escapes from the top of the atmosphere.
(Kasting et al., 1988).} 
\label{QUAL}
\end{figure}

The radiative model assumes a fixed surface temperature (which of course is not determined only by Solar flux but by all climatic factors entering the greenhouse effect) and the model is used to calculate the required net incident Solar flux, $F_S$, and the net outgoing infrared flux, $F_{IR}$, at the top of the atmosphere.\\
The \textit{effective Solar flux}, $S_{eff}$, is introduced as:
\begin{equation}
\label{SE}
S_{eff}= S/S_o=\frac{F_{IR}}{F_S}
\end{equation}
It is noteworthy that (\ref{SE}) is not a simply definition but the equality between the two last terms corresponds to two energy balances. The requirement is indeed the following: without atmosphere, but in the presence of \textit{albedo}, the ratio $S$ to $S_o$ is the fraction of incoming flux on the planet, needed to maintain a given effective surface temperature, $T_e$. It corresponds to the energy balance of Eq.(\ref{TB}), it means a given  ratio is requested between the emitted flux from Earth's surface (or planet's surface)  and the incident one $S_o$. This ratio must be equal also, when the atmosphere is present, to the ratio between the net outgoing IR flux from atmosphere and the net incident solar flux $F_S$ needed to obtain a given surface temperature $T_s$ (energy balances of Eq. (\ref{TBS})), both evaluated at the top of atmosphere  (see, also the qualitative picture of Fig.\ref{QUAL}) (Kasting et al., 1993, Kopparapu et al. 2013). \\
The planetary global \textit{albedo} is then obtained from the following expression\footnote{As we have seen, in general, $F_S$ corresponding to $T_s$, turns out to be less than the flux $S$ needed to obtain the same temperature as $T_e$, without greenhouse effect.} :
\begin{equation}
\label{AL}
A=1-\frac{4F_S}{S_o}
\end{equation}

Calculations based on this model were performed considering surface temperature, $T_s$, incrementaly growing from 220 K up to 2200 K (see, Kopparapu et al., (2013) which updated  the  Kasting et al. (1993) model).\\
From the energy balance (\ref{SE}) it immediately follows that: if the planet moves from 1 AU to d AU, $S$ and $S_{eff}$, scale according to:
\begin{equation}
\label{SD}
 S \Rightarrow S_{eff}\sim 1/d^2
\end{equation}
$S_o$ remaining fixed at Earth's orbit.\\
Moreover, if the luminosity of the central star in the planetary system changes, the following scaling law holds:
\begin{equation}
\label{SDL}
d= 1AU \left ( \frac{L / L_{\odot}}{S_{eff}} \right)^{0.5}
\end{equation}   

\subsubsection{Inner edge (IHZ)}
\begin{figure}[tb]
 \includegraphics[width=1.0\columnwidth,angle=00]{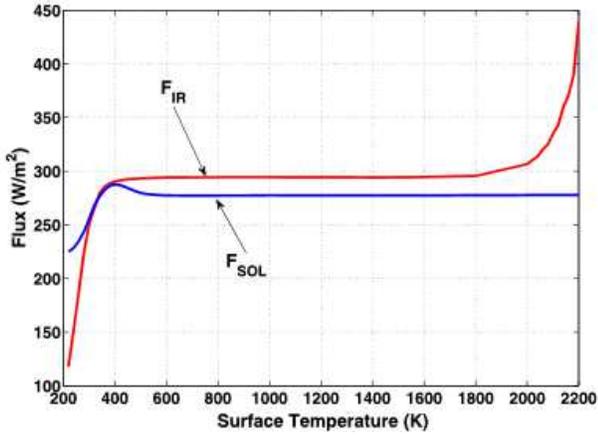}
\caption{Trends of fluxes ($F_{IR}$ and $F_{SOL}=F_S$) as function of surface temperature for the standard inner edge model (Kopparapu et al., 2013).} 
\label{TRENDS}
\end{figure}

\begin{figure}[tb]
 \includegraphics[width=1.0\columnwidth,angle=00]{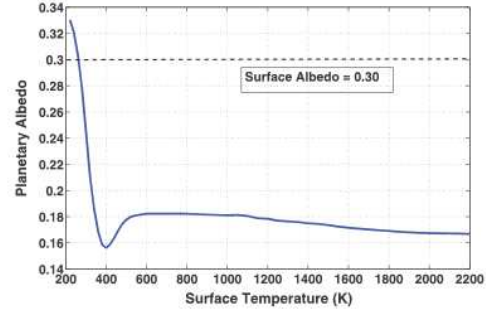}
\caption{As in Fig.\ref{TRENDS}, for planetary \textit{albedo} according to (\ref{AL}). The dashed line shows the assumed surface global \textit{albedo} (Kasting et al., 1993; Kopparapu et al., 2013c, Erratum).}
\label{ALB}
\end{figure}

The dependence of various physical quantities on surface temperature for the inner edge of HZ derived from the standard model are shown in 
Fig. \ref{TRENDS}, \ref{ALB}, \ref{SEFF} (Kopparapu et al., 2013).
Two limits for the inner edge can be calculated:\\
\begin{itemize}
\item the first one is the \textit{moist-greenhouse} or \textit{water-loss} limit (Fig.\ref{SEFF}). It occours at a surface temperature of 340 K ($\sim 67^oC$) when $S_{eff}=1.015$.Under these conditions the water vapor content in the stratosphere increases dramatically, by more than an order of magnitude. The orbital distance corresponding to the cloud-free water-loss limit is $d=1/S_{eff}^{0.5}= 0.99 AU$ for an Earth-like planet orbiting the Sun. That is considered a \textit{pessimistic inner edge distance.}
\item The second one is the \textit{runaway greenhouse} limit where the oceans evaporate entirely corresponding to $d= 0.97 AU$. The surface temperature at which it occours is about 1800 K ($\sim 527^oC$). This inner edge located closer to the star has to be considered a \textit{more optimistic inner edge.}
\end{itemize} 
 When the \textit{moist-greenhouse} limit is considered as inner bord, the criterium of \textit{water-loss } has to be taken into account, via \textit{photolysis} process in which the atmospheric $H_2O$ vapor molecules dissociate into $H^+$ and $OH^-$ ions, and the consequence of hydrogen escape (Kasting et al.,1993). The possible feedback of clouds is ignored. In the model of Kasting et al. (1993), the background atmosphere consists of 1 bar of $N_2$ and 300 parts per million by volume of $CO_2$, with no oxygen or ozone. In the model of Kopparapu et al. (2013) the background gas is instead made of 4 bar of $N_2$ and the total surface pressure is 6.5 bar.

\subsubsection{Outer edge (OHZ)}
To compute the outer edge, the Earth-like surface temperature was fixed at 273 K and the atmosphere was assumed to consist of 1 bar of $N_2$ while the $CO_2$ partial pressure was varied from 1 to 
35 bar (the saturation vapor pressure for $CO_2$ at that temperature) (Kopparapu et al., 2013). The working hypothesis is that atmospheric $CO_2$ would accumulate as the surface temperature decreases owing to the negative feedback provided by carbonate-silicate geochemical cycle.\\
To understand the existence of a \textit{maximum limit} for the greenhouse, we have to calculate the flux ratio ($S_{eff}$) required to mantain a global mean surface temperature of 273 K when $CO_2$ condensation, due to the increasing distance from Sun, occurs.The incident $F_{S}$ decreases monotonically with increasing $CO_2$ partial pressure (Fig.\ref{OUTER}) as a result of enhanced Rayleigh scattering. Correspondingly the planetary albedo increases (Fig.\ref{ALBC}). $F_{IR}$ decreases initially as $CO_2$ partial pressure grows. This is an indication of the \textit{greenhouse effect} of $CO_2$ (Fig.\ref{OUTER}). At about 10 bar, it asimptotically approaches a constant value as the atmosphere becomes optically thick at all IR wavelengths. So a minimum in $S_{eff}$ (Fig.\ref{SEFFMIN}) occurs at $CO_2$ partial pressure of $\sim 8 \ bar$ corresponding to a maximum distance $d= 1.67 AU$ which defines the \textit{maximum greenhouse} limit for the outer edge. It is very interesting to note that the end of the greenhouse effect is essentially due to conspiracy of two effects together: the decreasing of the $F_S$ owing to the increasing distance from Sun and the enhancement of Rayleigh scattering. At the same time, the $F_{IR}$  decreases while the reflecting power increases, as shown by the albedo trend (Fig.\ref{ALBC}). All that occurs in agreement with Kirchhoff's law: the emitted flux decreases as the absorbed one. Beyond this distance, $S_{eff}$ increases and the surface temperature of $0^oC$ is no more guaranteed by greenhouse effect.\\

\subsection{Venus and Mars as empirical optimistic limits}

We may notice that Venus, which is located at a distance $d=0.72 \ AU$, is too close to the Sun since the inner edge of the  HZ is equal to $0.97 \ AU$. However, Venus had surface features due to liquid water, which are absent for at least 1 Gyr. Then taking into account that the Sun at that time had about $\sim 92 \%$ of the present-day luminosity, Kasting et al. (1993) and Kopparapu et al. (2013) conclude that the IHZ might have been as low as $0.75 \ AU$ (recent Venus) suggesting that Venus was on the brink of a runaway greenhouse effect causing the disappear of liquid water. 
As for the inner edge model, a more optimistic empirical limit on the OHZ (Outer HZ) may be derived from the observation that the early Mars was warm enough for liquid water to flow on its surface when it was old of 3.8 Gyr. Even though the problem is still an open issue, taking into account that the Solar luminosity at that time was $\sim 75 \%$ of the present one, scaling the flux received by present Mars of that of Earth's to the early Mars,  Kasting et al. (1993) and Kopparapu et al. (2013) move the Mars distance from  $1.52 \ AU$ to $1.77 \ AU$ for the early Mars.\\

\subsection{Conservative and Optimistic HZ}
Following Kane et al. (2016), we summarize the limits of HZ obtained using the 1-D model considered:
 \begin{itemize}
 \item IHZ: \textit{moist and runaway greenhouse}, $0.99 \div 0.97 \ AU$; (\textit{recent Venus}: $0.75 \ AU$), 
 \item OHZ: \textit{maximum greenhouse}, $1.67 \ AU$; (\textit{early Mars}: $1.77 \ AU$).
 \end{itemize}
After the updated absorption coefficients for $CO_2$ and $H_2O$ (Kopparapu et al., 2013, 2014) the two IHZ have coalesced to $0.99 \ AU$. So the \textit{conservative theoretical HZ limits} and the corresponding values of the effective solar flux are, respectively:
$$d=(0.99 \ (0.95)-1.7) \ AU$$
$$S_{eff(Sun)}=(1.02 \ (1.11)-0.36)$$
whereas the \textit{optimistic empirical HZ limits} and the corresponding values of the effective solar flux are, respectively:
$$d=(0.75-1.8) \ AU.$$
$$S_{eff(Sun)}=(1.78-0.32)$$
To be noted that a better estimate for a theoretical IHZ is $0.95 \ AU$ (Leconte et al. (2013) by 3-D models, sect. 6.3.2). 
\subsubsection{Effect of clouds}

The HZ boundaries is strongly influenced by the presence of clouds ($H_2O$  and $CO_2$ clouds), even though quantitative statements are difficult. Their either warming or cooling effect depend on many parameters, e.g., their height, optical depth, particle size, and, most importantly, fractional cloud coverage. Calculations suggeste that $CO_2$ clouds should generally warm the climate rather than cool (Mischna et al., 2000). But on this topic the research is still in progress using 3-D models.\\
An additional atmospheric effect may be related to the presence of thick hazes
like that observed on Titan.
On one side they may cool the planet surface by about 20 K extending the inner
edge of the HZ but they may also grant UV shielding enhancing planetary
habitability (Arney et al., 2016).
 
\subsubsection{Extension of the HZ}

A significant broadening of the HZ, in particular the outer edge,  can occur adding additional greenhouse gases. Specifically if volcanism on terrestrial planets can maintain large amounts of molecular hydrogen,
a very light greenhouse gas (Ramirez \& Kaltenegger, 2017). Hydrogen retention
might be more efficient on super-Earths where the escape rates are lower
due to stronger gravity and the interior heat lasts longer favouring
vulcanism and $H_2$ outgassing. This mechanism, if active, may extend the
outer edge of the classical HZ by as much as $60\%$ (Ramirez \& Kaltenegger, 2017).\\
As climate models improve also the theoretical HZ limits evolve. 3-D climate models may include, e.g., factors such as variations in relative humidity and clouds which are impossible to estimate accurately in 1-D calculations. So by a 3-D study, Leconte et al. (2013) move the IHZ to at least $0.95 \ AU$. Also the inner edge of the HZ has been considered with 3-D climate model, for dry planets, sometimes called "Dune" planets. A (low-obliquity) planet of this kind would have water-rich oases near its poles. For such a planet with a Sun-like star, the IHZ could be as close as $0.77 \ AU$, about the empirical \textit{recent Venus} limit.\\
The inner edge of the HZ has been pushed closer to the host star ( until about 0.38 AU around a Solar-like star) if the \textit{greenhouse effect} is reduced (low relative humidity) and the surface \textit{albedo} is increased (Zsom et al., 2013). However this result has been criticized by Kasting et al. (2014). In conclusion, because it may take time for different models to reach consensus, we may refer to both conservative and optimistic estimates of the HZ from the 1-D model of Kopparapu et al. (2014) encompassing the limits from 3-D models (Kane et al., 2016, Fig.1).  

\begin{figure}[tb]
 \includegraphics[width=1.0\columnwidth,angle=00]{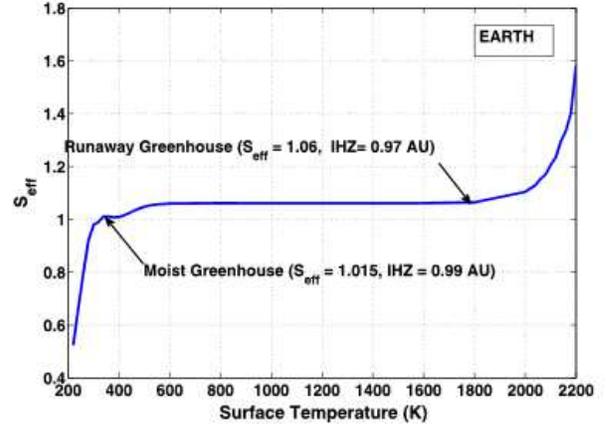}
\caption{As in Fig.\ref{TRENDS}, for effective solar flux, $S_{eff}$ with the two relevant inner limits, \textit{water-loss} (at a surface temperature of 340 K ($\sim 67^oC$)) and \textit{runaway greehouse} (at about 1800 K ($\sim 527^oC$)). (Kopparapu et al., 2013).} 
\label{SEFF}
\end{figure}

\begin{figure}[tb]
 \includegraphics[width=1.0\columnwidth,angle=00]{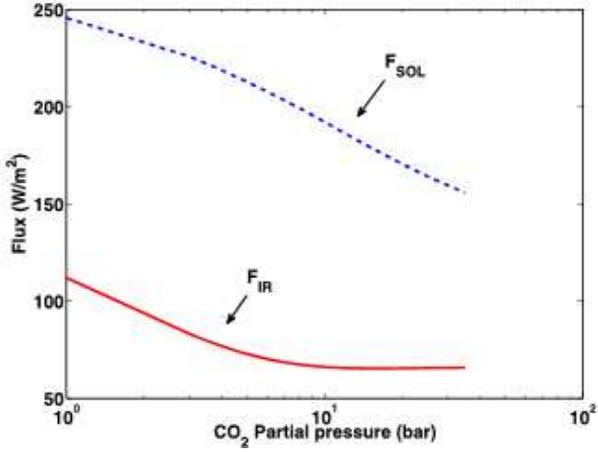}
\caption{Outer edge of HZ from climate model. Trends of Fluxes, like Fig.\ref{TRENDS}, as function of $CO_2$ partial pressure. The mean surface temperature is fixed at 273 K (Kopparapu et al., 2013).}  
\label{OUTER}
\end{figure}
\begin{figure}[tb]
 \includegraphics[width=1.0\columnwidth,angle=00]{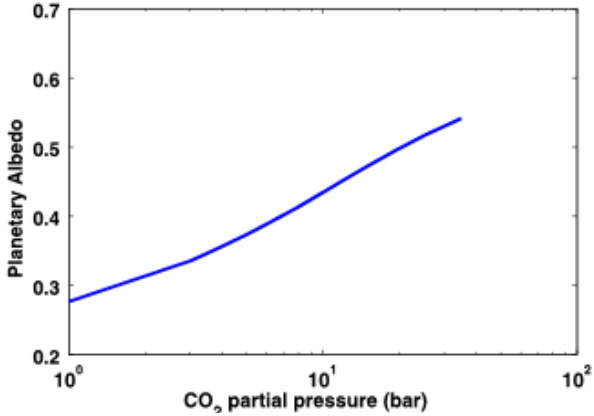}
\caption{As in Fig.\ref{OUTER}, for planetary \textit{albedo} (Kopparapu et al., 2013).}  
\label{ALBC}
\end{figure}
\begin{figure}[tb]
 \includegraphics[width=1.0\columnwidth,angle=00]{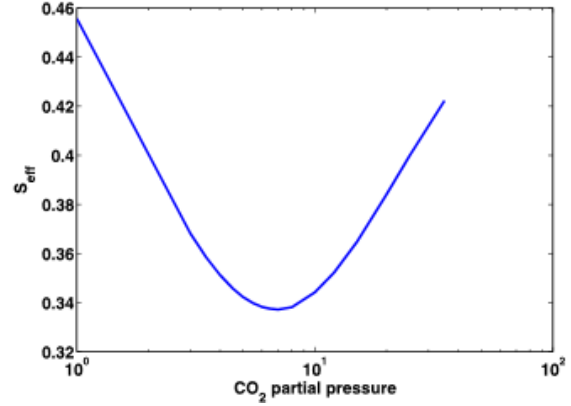}
\caption{As in Fig.\ref{OUTER}, for $S_{eff}$. Its minimum defines the \textit{maximum greenhouse limit} (Kopparapu et al., 2013).}
\label{SEFFMIN}
\end{figure}

\begin{figure}[tb]
 \includegraphics[width=0.90\columnwidth,angle=00]{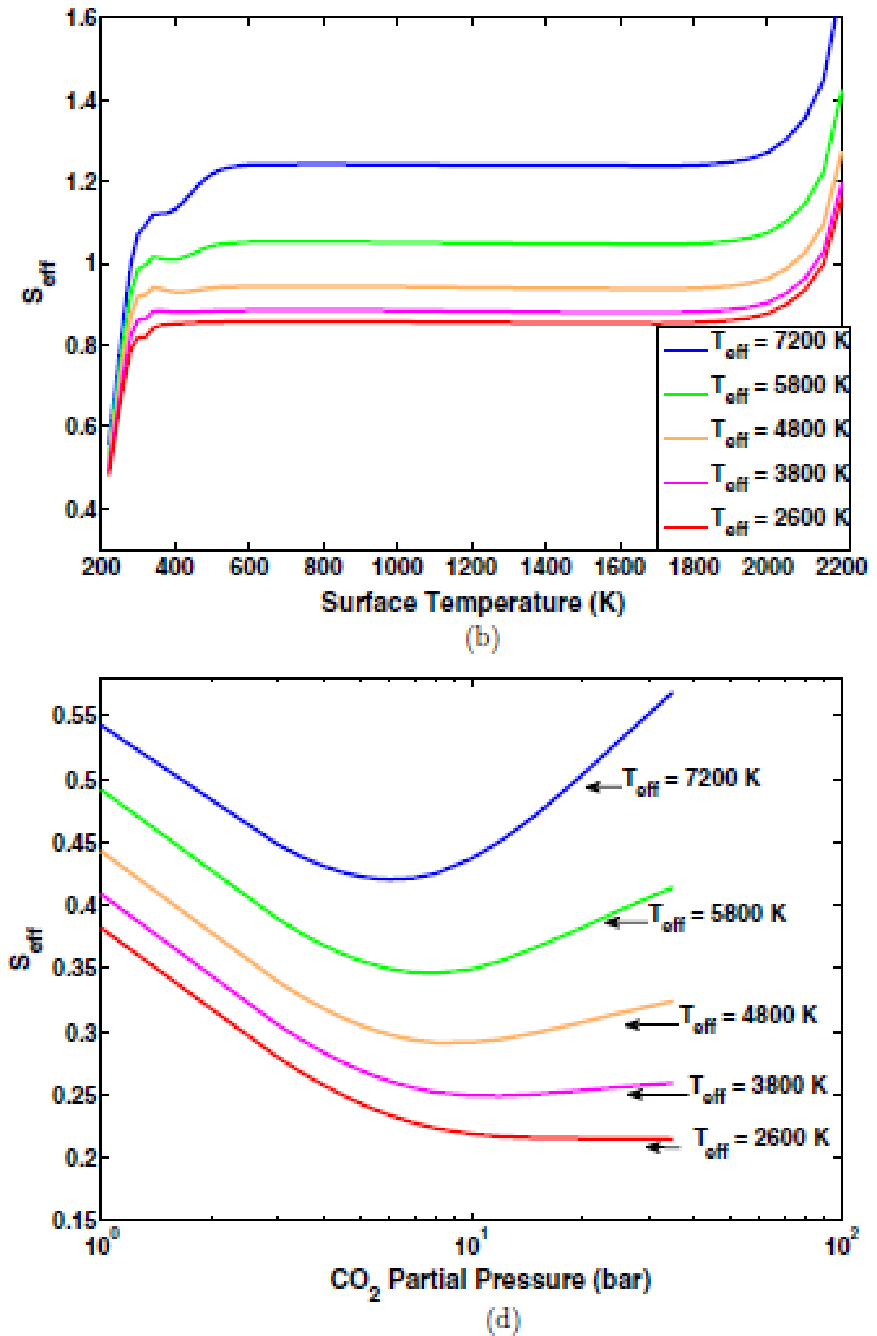}
\caption{Trends of $S_{eff}$ as function of surface temperature (for the inner edge, (b)) and of partial $CO_2$ pressure (for the outer one, 
(d)) for each $T_{eff}$ of the main- sequence stars hosting extra-Solar planetary systems, in the range $2600 \div 7200$.} 
\label{GENERAL}
\end{figure}



\section{Circumstellar Habitable Zone (CHZ) for extra-Solar systems}

\subsection{The method's generalization}

\begin{figure}[!ht]
\begin{center}
 \includegraphics[width=1.0\columnwidth,angle=00]{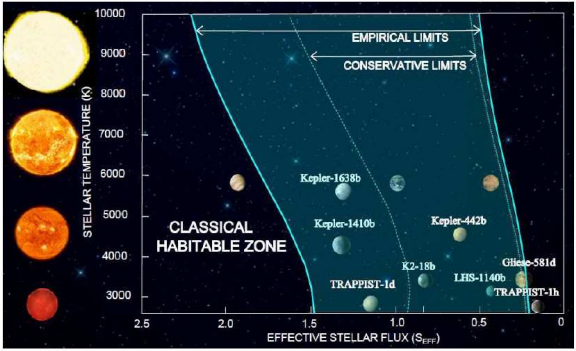}
\caption{Habitable zone with \textit{conservative} and \textit{empirical} limits in analogy with Solar system (sect.6.3), for a general planetary system around a main sequence star from 2600 to 10,000 K. Planets of TRAPPIST-1 systems, discussed in the text, are also shown (Ramirez, 2018).}
\label{HZ}
\end{center}
\end{figure}

Kopparapu et al. (2013) update the database used by Kasting et al. (1993) to build up their climatic model in the case in which at the center of the extra-Solar system there is a main-sequence star of spectral type F, G, K, M. The stellar effective temperature is in the range $2600 \ K \leq T_{eff} \leq 7200 \ K $. Classical HZ boundaries have been derived using the key Eq.\ref{SDL}, and the following relationship between HZ stellar fluxes ($S_{eff}$) reaching the top of atmosphere of an Earth-like planet and the stellar effective temperatures ($T_{eff}$) in the range considered:
\begin{equation}
\label{ESS}
S_{eff}= S_{eff\odot} + aT_* + bT_*^2 + cT_*^3 + dT_*^4
\end{equation}
where $T_*= T_{eff}-5780 \ K$ and $S_{eff\odot}$ is the the $S_{eff}$ value for a given HZ limit in our Solar system.  
The trends of $S_{eff}$ as function of surface temperature (for the inner edge) and of partial $CO_2$ pressure (for the outer one) are similar to those of the Sun for each $T_{eff}$ in the range $2600 \div 7200$ (Fig.\ref{GENERAL} vs. Figs.\ref{SEFF},\ref{SEFFMIN}). Recently the temperature range has been expanded to $10,000 \ K$ including then A-stars (Ramirez \& Kaltenegger, 2018). The quantities (a,b,c,d) are constant coefficients listed  in Table 1 of Ramirez (2018). This allows to define the Habitable Zone, with both limits \textit{conservative} and \textit{empirical} in analogy with Solar system (sect.6.3), for a general planetary system around a main sequence star as plotted in Fig.\ref{HZ}.
Some specific exemplifications are shown for Gliese-667C and TRAPPIST-1 systems.

\subsection{Application to the Gliese-667C and to the TRAPPIST-1 systems}
\begin{figure}[tb]
 \includegraphics[width=0.90\columnwidth,angle=00]{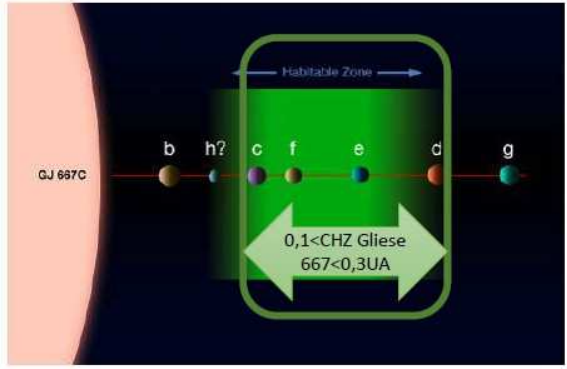}
\caption{Artistic picture of HZ in the Gliese-667C system (credit: ESO, adapted by Fecchio, 2016).
The two planets at the limit of HZ correspond to about the optimistic limits which, in the Solar system, are given by \textit{recent Venus} and \textit{early Mars} (Anglada-Escud\'e et al. 2013).} 
\label{ARTGLIE}
\end{figure}

The result for the optimistic HZ is shown  in the case of extra-Solar system with an M star (like Gliese-667C) at the center (artistic picture of Fig.\ref{ARTGLIE}). The characteristics of corresponding planets are listed in Anglada-Escud\'e et al., 2013. 


 \begin{figure}[tb]
 \includegraphics[width=0.90\columnwidth,angle=00]{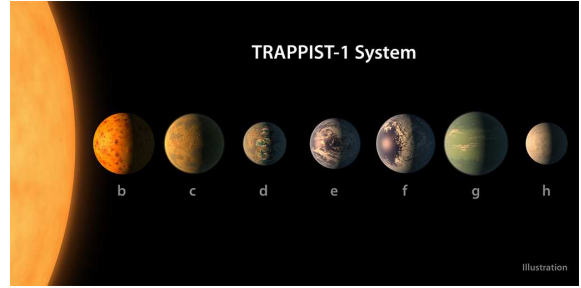}
\caption{Artistic picture of TRAPPIST-1 system ( credit: Nasa/JPL-Caltech-R. Hurt and T. Pyle). Trappist-1d and Trappist-1e might potentially be within the HZ for this system. } 
\label{ARTTRA}
\end{figure} 

The TRAPPIST-1 is an ultracool M dwarf with effective temperature of less than 2,700 K (Gillon et al., 2016). The seven planets of the system orbit the star with semi-major axes $ < 0.1$ AU and orbital periods of a few Earth days (Barr et al., 2018). Due to the star's low luminosity each planet has a moderate temperature, ranging from $\sim 160 \ K$ to 400 K. These low temperatures suggest that some might have solid surfaces composed of $H_2O$ ice and/or rock. Due to their non-zero orbital eccentricities tidal forces vary with time resulting in heating of their interiors by tidal dissipation. Tidal heating is an important energy source also in Ss for the satellites of the outer planets (e.g. for moons of Jupiter and Saturn).\\
Trappist-1d and Trappist-1e might potentially be habitable (Fig.\ref{ARTTRA}).Trappist-1d appears also in Fig.\ref{HZ} within the \textit{empirical limits} whereas Trappist-1h lies on the border of the same HZ. According to the models performed by Barr et al. (2018), planet $d$ could have a liquid water ocean at the surface and it might avoid the \textit{runaway greenhouse} state if its albedo is $\ge 0.3$. It seems also that the $e$ planet could harbor a thick ice mantle with likely liquid ocean underneath. To rule out a water-free structure, one needs to determine better the mass, to within $\sim 0.5 M_{\oplus}$. However, observations of TRAPPIST-1 showed frequent flares which constitute a serious threat for planetary atmospheres. Strong magnetic fields would be required to protect life from high energy radiation. Also the likelihood function for habitability calculated by Lingam \& Loeb (2017c) for planet $e$ turns out to be at least one to two orders of magnitude lower than in the case of Earth.

\section{General remarks}
We wish to point out that the aim of previous sections  was to consider the life as done, similar to what we know on Earth and of the kind considered in sect.2. Under this assumption, our goal was to explore some conditions for harboring and preserving it, and this determines the HZ limits, without any considerations about the origin of life. The introduction of climate model and its link with dynamics (sects. 5) is then consistent with our aim on this planetary scale, but some topics related to chapter on prebiotic chemistry  and astronomy have been intentionally skipped. Nevertheless, let us mention something more of this very wide excluded perspective, at least for what is inherent to some aspects here considered. Ad example, in sect.2 one of the necessary request ( $\delta$)) for a site suitable for life was to allow formation of long molecular biological chains, with the implicit reference to polymer assembly. A very interesting mechanism has been applied by Lathe (2004, 2005) to biomolecules capable of association/polymerization at high salt concentration and of dissociation at low salinity. Nucleic acids like DNA meet these criteria. Lathe postulated that fast tidal cycling may be connected with the above request and also with the origin of life. At ocean shores tides may provide association/assembly on drying while charge repulsion on tidal dilution drive dissociation. If this mechanism really works that might suggest constraints on the evolution of extra-terrestrial life as Lingam \& Loeb (2017a) claimed. They  analyze the conditions under which tides may also exert a significant positive influence on abiogenesis and biological rhythms. Indeed most biological organisms on Earth have evolved internal timing mechanisms even though it is not clear if these kind of biological clocks are induced into life or they instead are key processes to obtain life. \\

\subsection{Fast tidal cycling at the beginning of life}

Whatever the answer may be it turns out interesting the comparison between what happens on Earth at origin of life and what is to be expected in some extra-Solar systems. It seems that the rotation period of the Earth was $\sim 5 \ h$ at the time of Moon's formation (Brush, 1986). This rapid rotation might be due to the impact event that produced Moon (about 5 Gyrs ago). Moreover at the origin of life (about 3.9 Gyrs ago) the Moon was probably much closer to the Earth than present (perhaps only 200,000 km away) (Brush, 1986). As consequence, tidal forces were stronger than today and they also had a shorter period. According to Lathe (2004), large local fluctuations in salinity concentration could have provided effects on prebiotic precursor molecules and then also on the origin of life. Lingam \& Loeb (2017a) argue about the possibility that something similar is also to be expected in planetary systems around low-mass stars (e.g., around the M star Gliese-667C, or TRAPPIST-1, an ultracool M dwarf  (see sect. 7.2)).


\section{Moving from CHZ toward GHZ and COSHZ}
It is to noteworthy that the criteria adopted for establishing HZ run around only two necessary conditions (connected one to the other) in order to host life: to reach a reasonable surface temperature and to allow the presence of liquid water. These two conditions are met depending on the distance from Sun and on all the climatic factors which contribute to the greenhouse effect. However, we are dealing with necessary but not sufficient conditions. If we compare the requirements for the Earth to harbour life (sect.4), the conclusion is that some extra-Solar planets which lay within the habitable strip may not be really habitable (see also the remarks done in sect. 7.2, for the planets of TRAPPIST-1 system).\\ Moreover, in the next sections, we emphasize that to be sure a candidate lies within the HZ doesn't avoid the question: \textit{in what part of the Galaxy is it located?}. 
Moving from the planetary scale (CHZ), the problem of the metal amount needed to form a planet like Earth, the threats of supernovae explosions and possible comet injections due to tidal Galactic effects, will force us to meet the wider chapter of Galaxy formation and evolution, introducing the Galactic Habitable Zone (GHZ). From that, in turn the horizon gets larger to the Cosmology because an interplay exists between some crucial factors constraining the main features of the Universe and its evolution with life. The set of these fine tuned links defines, in our view, the cosmological habitability (COSH) (sect.15).\\

\section{Galactic Habitable Zone (GHZ)}
In order to set some limits within the Galaxy for the birth and development of life, we first consider locations where it is possible to form Earth-like planets. The corresponding space-time location at Galaxy scale, suitable for life is constrained by the \textit{metallicity amount} that the chemical-dynamical evolution of the Galaxy is able to yield.\\
 Moreover to satisfy the requests $\gamma)$ and $\delta)$ of sect.1, we have to protect life from supernovae explosions and injection of comets (sect.12 and 14.4).\\ In other words we at least have to take into account:\\
\begin{itemize}
\item the metal amount; 
\item the distance from Galactic center;
\item the supernovae explosions;
\item the Galactic tidal effects on an analog of our Oort's cloud, which might trigger a cometary bombardment towards the inner part of the planetary system.
\item the time needed to develop a complex Life (already considered in sect.2).
\end{itemize}
A short historical synthesis about the limits of GHZ due to different authors, is given in Tab.\ref{tbl:1}.\\
Introduced for the first time by Gonzalez et al. (2001) the GHZ  was essentially based on the metallicity gradient on the Galaxy disk. One of the more important factors is indeed the metallicity of the interstellar matter out of which a planetary system forms which determines the masses of terrestrial planets. The assumption was that lower metallicity
led to smaller planets. So the upper limit for GHZ was estimated, very approximately, the Galactic location where the metallicity is, at least, half that of the Sun. The key issue was the production of terrestrial planets with a mass able to sustain plate tectonics (preliminary results about its dependence on the mass has been later on given by Noack \& Breuer, 2011). The inner limit of this newly defined GHZ was set by high energy events. But the boundaries of GHZ remained not well defined.

\begin{table}[ht]
 
\caption{Limits of the Galactic Habitability Zone given by different authors. The time satisfies the request for complex life (see, next sections).}
\label{tbl:1}
\begin{center}
\begin{tabular}{lrr}
\hline 
Time (Gyr) &Author& Limits\tabularnewline
\hline 
4-4,5 & Gonzalez et al. (2001) & 4.5-11.5 kpc\tabularnewline
\hline 
\hline 
4,5 & Lineweaver et al. (2004) & 4-11 kpc\tabularnewline
\hline 
4,0 & Lineweaver et al. (2004) & 4-10 kpc\tabularnewline
\hline 
$\sim$5,0 & Prantzos (2006-2008) & 5-16 kpc\tabularnewline
\hline 
 \end{tabular}
      \end{center}
\end{table}


\subsection{Contribution of Lineweaver et al. (2004)}
For defining better the GHZ they  investigate the following items:\\

a)- the estimation of how many stars are available on the Galactic disk for hosting a planetary system. To this aim, the space-time trend for the \textit{star formation rate} (SFR) has to be known.\\

b)- the lower and upper limits for the amount of the heavy elements which are necessary to build up a terrestrial planet.
Too little metallicity means indeed a lack of necessary material to form a planet as Earth. Conversely too much of it would cause formation of a large number of massive planets, as we will see in sect. 11.2.1\\

\begin{figure}[!]
\includegraphics[width=0.7\columnwidth,angle=00]{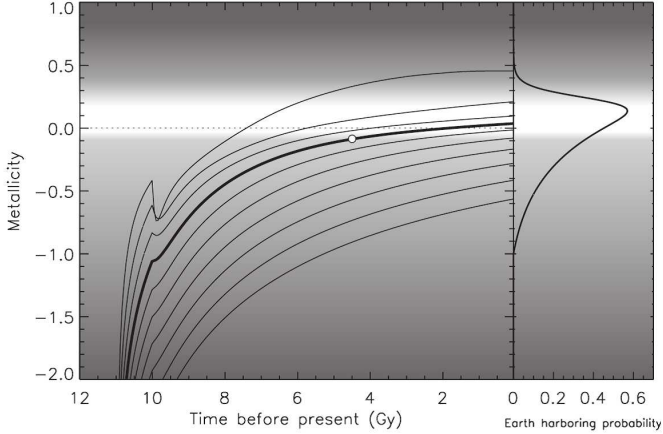}
\caption{Spatial distribution of metallicity at different Galactocentric distances from 2.5 kpc (upper curve) to 20.5 kpc (lower curve) with increments of 2 kpc, is shown as funtion of time. The white dot indicates the Sun's time of formation at Galactocentric distance of 8.5 kpc. On the right, the probability to harbour terrestrial planets as function of star host metallicity (based on chemical-dynamic model of Fenner and Gibson, (2003); see, Lineweaver et al., 2004).}
\label{ATempo}
\end{figure}

To give the probability, $P_{metals}$, to host terrestrial planets, they need to know the space-time distribution of metals within the Galaxy. The answer is given in Fig.\ref{ATempo}, obtained by a chemical-dynamic model (Fenner and Gibson, 2003).\\

Moreover, they consider:\\

c)- the time needed for developing biological evolution. According to biological time clock of Fig.\ref{OROLOGIO} (item $\beta)$), a typical time of $4\pm 1 Gyr$ is chosen. This constraint for complex life has been modeled as a probability, $P_{evol}(t)$, defined as cumulative integral of a gaussian distribution caracterized by a mean value of $4 \ Gyr$ with a dispersion of $1 \ Gyr$.\\

As additional constraint, the necessity:\\

d)- to avoid the supernovae whose high energy radiations may be indeed very dangerous for life. The probability that complex life has to survive to this catastrophic event is defined as, $P_{SN}=0.5 \ \xi (r,t)$, where $\xi (r,t)$ is a danger factor depending on Galactocentric distance $r$ and on the time $t$ of star formation.

\begin{figure}[!]
\includegraphics[width=1.0\columnwidth,angle=00]{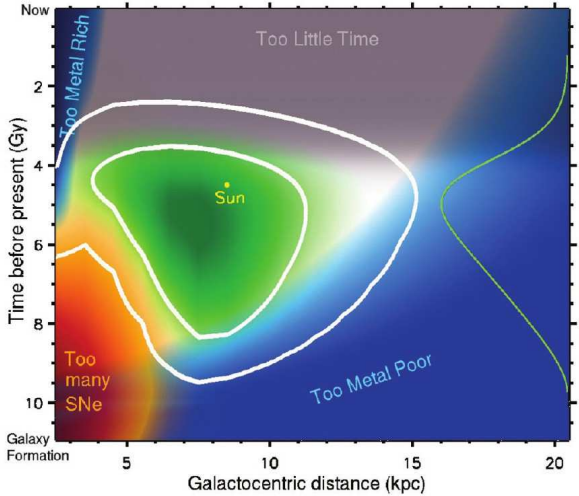}
\caption{GHZ (in green) on Galaxy disk in the plot Galactocentric distance vs. \textit{lookback} time, obtained by the requests based on the probability to harbour stars with suitable metallicity (blue region), with enough time for complex life evolution (gray region), and with freedom from life-extinguishing supernovae explosions (red region). The white contours encompass $68\%$ (inner) and $95\%$ (outer) of $P_{GHZ}$, including then the stars having the highest potential to be harboring complex life. The green line (on the right) shows the age distribution of the probability  needed by complex life (see, text) (Lineweaver et al., 2004).} 
\label{IM1} 
\end{figure}

At the end the relative number of potentially suitable planetary systems as a function of space and time, is given by:
\begin{equation}
\label{proeq}
P_{GHZ}(r,t)=SFR \cdot P_{metals}\cdot P_{evol}\cdot P_{SN}
\end{equation} 
which means the composite probability  of obtaining on Galaxy disk a site favourable for the development of complex life. 
The two factors $P_{SN}$ and $SFR$ are not independent (indeed as SFR decreases, $P_{SN}$ increases). 
The GHZ is identified as the region in the disk bounded by the limit of $68\%$ (until $95\%$) of $P_{GHZ}(r,t)$ (Fig.\ref{IM1}).
The above constraints define a zone centered  around $8 \ kpc$ from the Galactic center growing with from $8 \ Gyr$ to $4 \ Gyr$ before present (\textit{lookback time}) (Fig.\ref{IM1}). By integrating $P_{GHZ}(r,t)$ over $r$ we obtain the age probability distribution of complex life along the whole disk (on the right of Fig.\ref{IM1}). This tells us that $\sim 75\%$ of the stars that could harbor complex life in the Galaxy are older than the Sun of about $1 \ Gyr$. \\

\begin{figure}[!]
\includegraphics[width=1.0\columnwidth,angle=00]{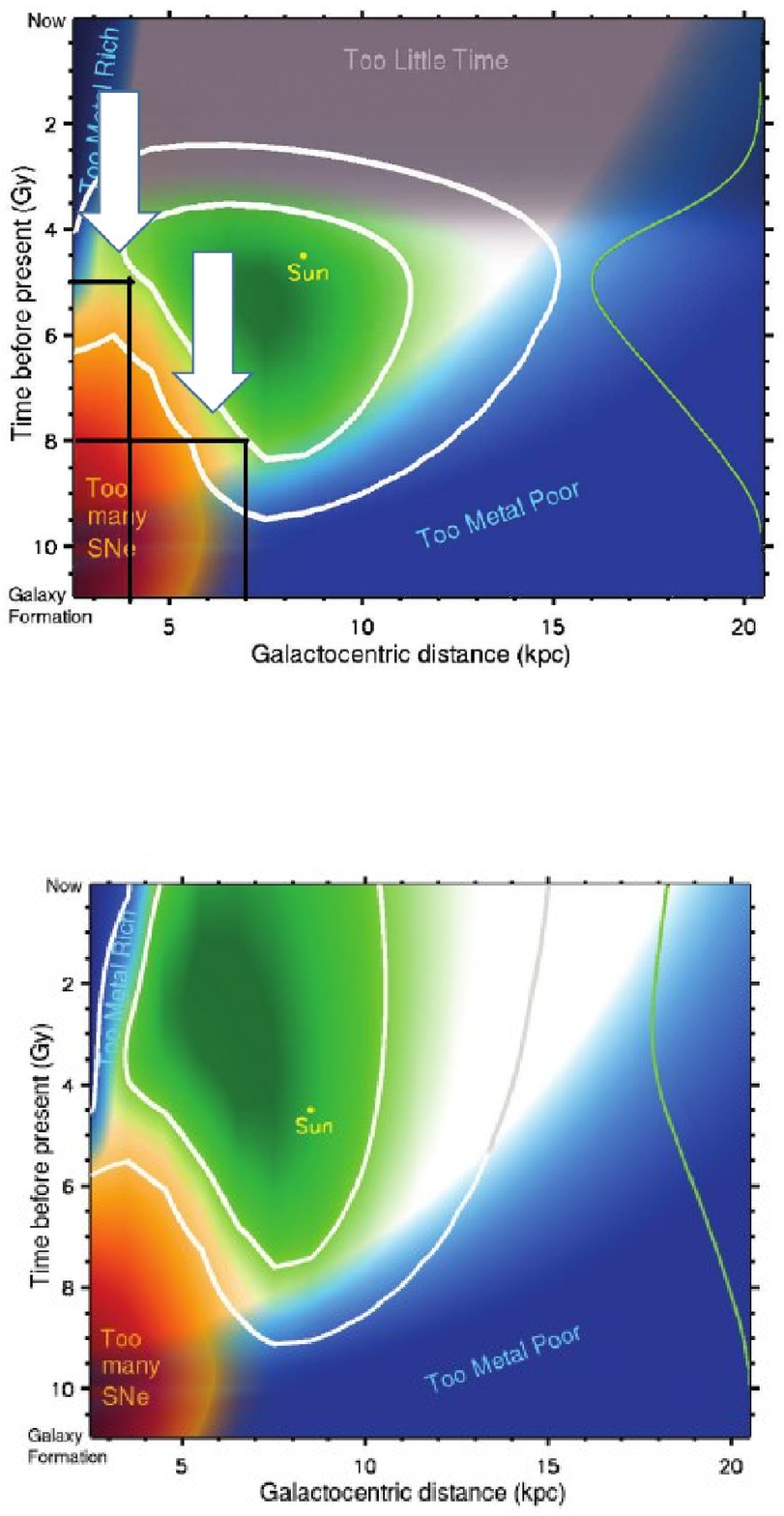}
\caption{GHZ as in Fig.\ref{IM1}; see text for the meaning of the white arrows in the upper panel (Lineweaver et al., 2004, adapted by Fecchio, 2016). Variation of GHZ when the upper limit for the time to develop complex life is relaxed. Again, the green line on the right is the resulting age probability distribution of life along the whole disk (lower panel, Lineweaver et al., 2004).}  
\label{IM4}
\end{figure}

It is remarkable to note that (Fig.\ref{IM4}):
\begin {itemize}
\item At the beginning of Galactic history the high star formation rate in the inner part of the disk (under $7 \ kpc$ and at about $8 \ Gyr$, white arrow on the right) was able to produce heavy elements in order to form terrestrial planets but, at the same time, it caused, an unacceptable so high emissions from supernovae able to inhibit the rise of life during some Gyr (e.g. at $4 \ kpc$,  about $3 \ Gyr$ further must be spent, white arrow on the left).
\item the Galactic bulge, which extends until about $3.5 \ kpc$, turns out to be not an environment suitable for life because too crowded with stars and then potentially able to produce strong supernovae emissions and star encounters.
\item As soon as the time constraint for reaching the evolution of complex life is relaxed, the Habitability Zone ($68\%$) expands upwards, as shown in the lower picture of Fig.\ref{IM4}). 
\end{itemize}
In conclusion, from Lineweaver's et al. (2004) contribution two slightly different upper limits for the GHZ are obtained which depend on the different time needed to produce complex life (Tab.1): $11 \ kpc$, in the case of $4.5 \ Gyr$ or $10 \ kpc$, in the case of $4.0 \ Gyr$. By relaxing the time constraint for developing complex life, the GHZ nowadays is in the range of about $4.5\div 10 \ kpc$ (Fig.\ref{IM4}, lower panel).

\subsection{Contribution of Prantzos (2006, 2008)}
Prantzos is strongly critical about the limits given for GHZ by Lineweaver et al. (2004) being also very doubtful on the possibility to define them. In his opinion some assumptions which appear in Eq.\ref{proeq} are far from being unequivocally defined. He critically reviewed them. 

\begin{figure}[!]
\includegraphics[width=1.0\columnwidth,angle=00]{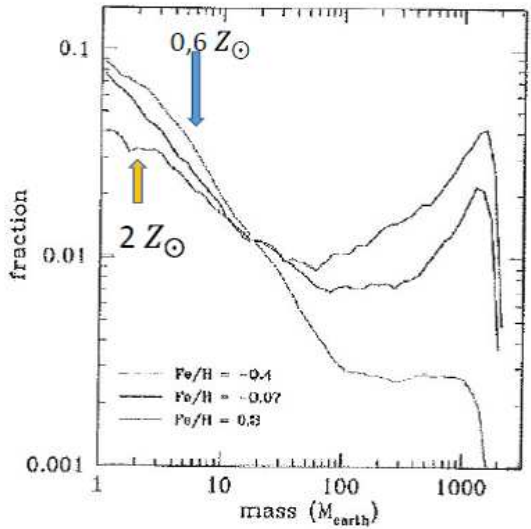}
\caption{Fraction of planets with different masses (in $M_{\oplus}$) which may form around a star of $1 \ M_{\odot}$  as function of three different metallicity values for star (and corresponding proto-planetary disk) in the range ($0.6 \ Z_{\odot}  \div \ 2 \ Z_{\odot}$) . Increasing metallicity favours also the formation of giant planets (simulations of Mordasini et al., 2006).}
\label{pranz_delta_z}
\end{figure}

\subsubsection{Metal amount}
\begin{figure}[!]
\includegraphics[width=1.0\columnwidth,angle=00]{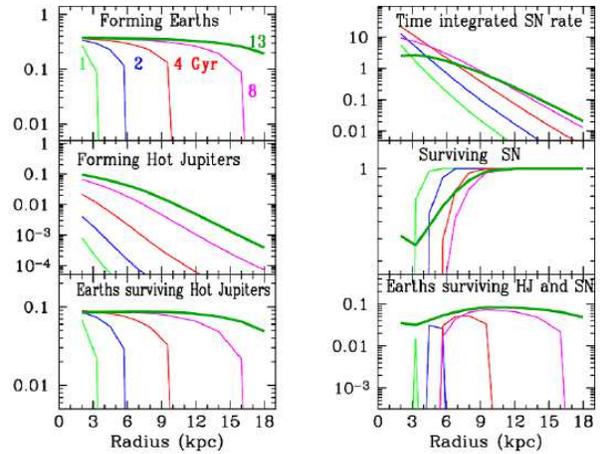}
\caption{Probabilities of various events as a function of Galactocenric distance at five
different epochs of the Milky Way evolution: 1, 2, 4, 8 and 13 Gyr, respectively (the
latter is displayed with a thick curve (green) in all figures). The probabilities of forming
Earths is given in the upper left panel and of forming Hot Jupiters in the middle left one. To form Earths but also to survive the "Hot Jupiters" in the lower left panel 
, while
the probability of life bearing planets surviving SN explosions is giving in the right middle panel. Finally, the overall probability for Earth-like planets surviving HJ and SN is shown on bottom right. Initially the GHZ 
defines a ring in the MW disk, progressively migrating outwards;
that "probability ring" is quite narrow at early times, but fairly extended today
and peaking at about 10 kpc (Prantzos, (2006-2008)).}
\label{IM5} 
\end{figure}

To determine the metal amount two opposite requests have to be taken into account: the first is to have enough metals to form planets similar to the Earth, the other one is to avoid that the increasing metallicity produces also too many giant planets (so called "Hot Jupiters") 
whose masses are comparable (or superior) to that of Jupiter. In Fig. \ref{pranz_delta_z}, the distribution function of planet vs mass (in Earth mass) is illustrated for three metallicity abundances of the central star and then of the proto-planetary disc. In principle, the gaseous giants form at a distance of several AU from their stars where the gas is available for accretion onto a rapidly growing rocky/icy core. By tidal interaction with the proto-planetary disk, they then lose angular momentum causing their inwards migration. During migration they may dynamically destabilize and eject from the system smaller planets which might host life.\\ 
In this scenario, we have to take into account not only the probability to form Earth-like planets but also to estimate the probability that they have to survive in presence of migrating giant planets (Fig.\ref{IM5}). The resulting probability
of having Earths surviving Hot Jupiters (HJ) is given as a function of Galactocentric distance at five different epochs of the Milky Way evolution: 1,2,4,8 and 13 Gyr, time-frames are shown on the last panel on the left of (Fig.\ref{IM5}).
 
In addition, also the probability of surviving supernovae explosions has to be re-considered. Prantzos' opinion is that: if on the base of destructive effect of one SN explosion lies its capability to produce energetic particle/flux irradiation in order to induce a significant number of \textit{gene mutations}, then one may calculate a lower limit for the distance (D) of the SN able to produce such  fluxes/irradiations  and an estimate of the Galactic SN frequency occurring at this distance.
According to Gehrels et al. (2003), significant biological effects due to ozone depletion, i.e. doubling of the UV flux on Earth's surface, may arise for SN explosions closer than $D\sim 8 pc$ with a frequency $f\sim 1 Gyr^{-1}$ (evaluated by Prantzos). However Prantzos' opinion is that it appears hard to draw any quantitative conclusions about the probability of \textit{definitive sterilisation of a habitable planet} because, e.g. the considerations done are relative to all land animals whereas marine life could probably survive to a large extent (since UV is absorbed from a couple of meters of water). The overall probability to have Earths surviving HJ and SN at different Galactocentric distance and at different epochs is drawn in the last panel on the right of (Fig.\ref{IM5}). 

\subsubsection{Prantzos vs. Lineweaver et al. 2004}

Despite his perplexities, Prantzos gives the evolution of GHZ during the cosmological (universal) time, from 1 to 13 Gyr which is shown in final panel on the right of Fig. \ref{IM5} . At 2 Gyr it appears as a very narrow ring of about $(4\div 6) \ kpc$ (in blu) progressively growing at 4 Gyr to the extension of about $(5\div 10) \ kpc$ (in red), reaching the dimension of about $(5\div 16) \ kpc$ at 8 Gyr (magenta), including, practically now (13 Gyr), about the whole extension of the Galactic disk (thick curve in green). By comparing the results of Prantzos and Lineweaver's et al. (2004) and considering that they take into account the \textit{lookback time} whereas Prantzos does not include the time needed for complex life, we can conclude that their strip at 4 Gyr may be assimilated with that of Prantzos at 8 Gyr (because it means allowing about 5 Gyr for complex life also in his approach), the Prantzos' GHZ turns out to be only slightly larger than that of Lineweaver et al. (2004), both having an increasing trend towards present. As soon as the term related to the time for developing complex life is relaxed ( it means to consider for both the GHZ at the present) the difference between the two approaches are more relevant: for Prantzos the GHZ extension covers the whole disk with a maximum around $10 \ kpc$, for Lineweaver's et al. (2004) the extension is only of \  $(4.5\div 10) \ kpc$.\\
We have also to point out that Prantzos (2006-2008) built up also probability isocontours of having an Earth-like planet surviving SN explosions in the time-distance plane. But because that refers to a planetary system around \textit{one star} at a given time-space position, he  multiplied this probability with the corresponding surface density of stars in order to obtain the probability of having life \textit{per unit volume} in a given position. All that leads to the following paradoxial conclusion: despite the high risk from SN early on in the inner disk, this same place later becomes relatively "hospitable" because of large star density in it.\\
To similar conclusion arrive also Gowanlock et al. (2011). In such way the Solar neighborhood in the outer disk appears to be not privileged to host life whereas the more suitable environment for life appears to be in the inner disk with a probability peak at about $2.5 \ kpc$.

\subsection{A criticism}
Against these two paradoxial conclusions which might leads to the drastic Prantzos' thinking that: \textit{the concept of a GHZ may have little or no significance at all}, our criticism is the following: in his reasonings Prantzos takes into account the SN frequency which already needs implicitly the inclusion of SFR (conversely  included explicitly by Lineweaver et al. (2004) in their probability Eq.(\ref{proeq})). Then probably Prantzos and Gowanlock's et al. last conclusions at the end of previous sect., are affected by considering two times the weight of the SFR.\\
That is the reason why we will continue to refer as a relevant proposal to the GHZ given by Lineweaver et al. (2004) improving it by adding the 
tidal Galactic effects on comets of Oort's Cloud or analog in extrasolar systems.   
   
\section{Galactic planar tides on the comets of Oort Cloud and analogs}

The Oort's Cloud (Oort, 1950) is the natural reservoir for long period comets of our Solar System: it is an outer shell structure roughly placed between $20000 \ AU \ (\sim 0.3 \ ly$) and $150000 \ AU \ (\sim 2.2 \ ly$) from Sun.

The comets represent the remnants of the formation of our planetary system: they have been driven in the Oort cloud through a scattering process induced by proto-planets combinated with a Galactic tidal torque effect at the beginning of the history of the Solar System (Tremaine, 1993). The existence of other comet clouds in different planetary systems spread out in the Galaxy is suggested by recent evidences of dusty excess on debris disks in exoplanetary systems observed by Spitzer at 70 $\mu$m (Greaves \& Wyatt, 2010). 
The Galactic tide may re-inject the Oort Cloud's comets towards the inner part of the planetary system producing  a comet flux with possible impacts on it and then with consequences on the development of life.

\begin{figure}[!]
\includegraphics[width=0.9\columnwidth,angle=0]{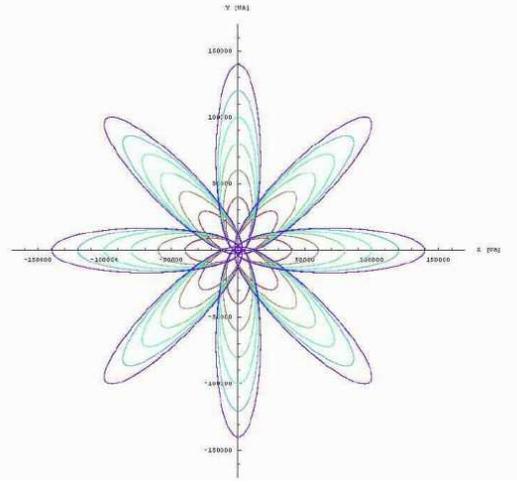}
\caption{Distribution of comet orbits in an inertial frame on the Galactic plane with the same perihelion and increasing aphelions at different Galactic longitudes (see text) before switching on the Galactic tides (Masi et al., 2003; Masi et al., 2009). The initial not perturbed orbits here considered are similar to those taken into account by De Biasi et al. (2015).}
\label{MASI_1}
\end{figure}

\begin{figure}[!]
\includegraphics[width=0.9\columnwidth,angle=0]{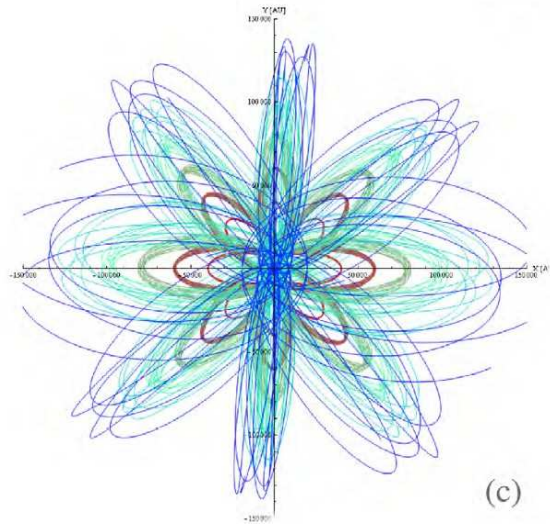}
\caption{As soon as Galactic tides are on, they affect the comet's orbits of Fig.\ref{MASI_1}, as stronger as the Solar system is placed closer to the Galactic center ( c), at $4$ \ kpc) (Masi et al., 2003; Masi et al., 2009).}
\label{MASI_2_3}
\end{figure}

The effects of the Galactic tides on the Oort Cloud's comets were studied by several authors (e.g., Heisler \& Tremaine, 1986; Matese \& Whitman, 1992; Fouchard et al., 2006). Usually they are separated into radial, transverse and orthogonal (to the disk plane) components. At solar distance from the Galactic center the orthogonal tide has been estimated to be about ten times more effective than the radial one, in perturbing the Oort Cloud's comets into the Solar System  (Heisler \& Tremaine, 1986; Morbidelli, 2005). At this distance, it appears to be the dominant perturber today and also over the future long term (Heisler, 1990). 
In spite of that, the contribution of planar tides (radial and transverse) becomes non negligible when the distance of the parent star, with respect to the galactic center, changes, as shown by Masi et al.(2009) for the Oort cloud comets and by Veras \& Evans (2013) in their analysis on dynamics for extrasolar planets. In particular the planar component of the tide may cause a great change if the central star decreases its distance and/or its mass with respect to the present solar one. 
Moreover they will certainly be non negligible in presence of spiral arm perturbations, as also Kaib et al. (2011) have already pointed out and it has been confirmed in the preliminary results of De Biasi (2014).\\
An interesting contribution to this problem comes out by the work of De Biasi et al. (2015). They consider only the effects of Galactic planar tides but at two different distances from the Galactic center: $4$ \ kpc and $8$ \ kpc, assuming the last one as the current solar position. That is performed in the perspective that the Sun may have migrated through the Galactic disk from an inner zone to the current one (Kaib et al., 2011) (sect.14). This means that the solar system could have been subjected to tidal perturbations stronger than those which characterize the present environment. The plastic impression has already been given in the paper of Masi et al. (2009). Indeed in their Fig.\ref{MASI_2_3}, the effect of increasing total Galactic tidal perturbation moving toward the center of the Galaxy, clearly appears.
In an inertial system of reference they studied the behavior of $48$ test particles (simulating comets) with a common perihelion $q=2000 \ AU$ and an increasing aphelion $a=(40,60,80,100,120, 140) \cdot 10^3 \ AU$, repeating each group at the following Galactic longitudes: $(0^o, 45^o, 90^o, 135^o, 180^o, 225^o, 270^o, 315^o)$ (Fig.\ref{MASI_1}).\\

\subsection{The contribution of dynamical Galaxy components to the tidal effects on comets}
The main contributions to Galactic tide have their origin from the gravitational potential of the bulge $\Phi_{BG}(r)$, of the disk $\Phi_D(R,z)$ and of the dark matter halo $\Phi_{DH}(r)$ respectively given by (De Biasi et al., 2015):
\begin{equation}
\Phi_{BG}(r)= -\frac{G M_{BG}}{\sqrt{r^2+r_c^2}}
\end{equation}
where $M_{BG}$ is the bulge total mass, $r_c$ is the \textit{core radius} (Flynn et al., 1996) and $G$ the gravitational constant, assuming as a good approximation the spherical model of the \textit{Plummer sphere} (Binney \& Tremaine, 2008).\\
For the \textit{thin} axisymmetric disk the infinitely thin Freeman's model (Freeman, 1970) appears as main reference. But, in this context it is 
affected by some difficulties which may be overcame combining instead three Miyamoto-Nagai disks (Miyamoto \& Nagai, 1975) of differing scale lengths and masses:
    \begin{equation}\label{eq:MN}
    \Phi_{MN}(R, z)= - \sum^3_{n=1}\frac{G M_n}{\sqrt{R^2+\left[a_n+\sqrt{b^2+z^2}\right]^2}}.
    \end{equation}
    The parameter $b$ is related to the disk scale height, $a_n$ to the disk scale lengths and $M_n$ are the masses of the three disk combined components.
It needs to stress that the potential trend given by Freeman's model results in fair agreement with that of (\ref{eq:MN}) at the mid-plane ($z=0$). 

The Galactic mass distribution of the DM halo is still uncertain and there are several models in the literature.
Instead of classical Navarro-Frenk-White (NFW; Navarro et al., 1997) density profile, the modified pseudo-isothermal (MPI) profile (Spano et al., 2008) is considered. That has been introduced to obtain the best fits of rotation curves in spirals and for LSB (low surface brightness galaxies), probably DM dominated. The corresponding gravitational potential is given by:
      \begin{eqnarray}
			\nonumber
      \Phi_{DH}^{MPI}(r)=-4\pi G \rho_0 r^2_H\times\\
			\left(\frac{\mbox{ln}\left(r/r_H+\sqrt{1+\left(\frac{r}{r_H}\right)^2}\right)}{r/r_H}
			-\frac{1}{\sqrt{1+\left(\frac{R_{vir}}{r_H}\right)^2}}\right),
      \end{eqnarray}
with $\rho_0$ the central dark halo density.
The values of the constants $a_n$, $b$, $M_n$, according with Flynn et al.(1996), and the parameter values for each Galactic components are listed in Tab.1 of De Biasi et al. (2015).        







      
\subsubsection{Circular velocity and tide contributions from Galaxy components}
\begin{figure}[htbp]
      \centering
      \resizebox{\hsize}{!}
      {\includegraphics{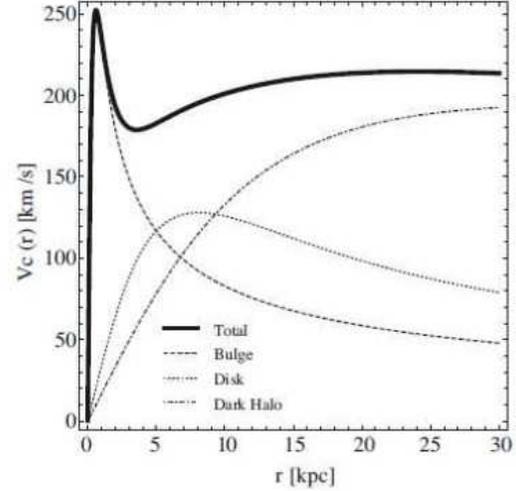}}
\caption{The different contributions to the circular velocity vs the distance from the center are here plotted for Galaxy assuming a MPI density profile for DM halo (Spano et al., 2008) (De Biasi et al., 2015).}  
\label{rotc}
\end{figure}

The contributions to the total circular velocity trend vs. $r$ from different Galaxy components are plotted in Fig. \ref{rotc}.
The values for solar environment are inside the measured error bar for these distances (Klypin et al., 2002)

As we will see in the next sect., one relevant not expected result is the following: the contribution to the circular velocity coming from a dynamical Galactic component, enters in determinant way also the description of tide coming from the same component as soon as it is done in Hill's approximation (Eq.\ref{ratid}). Because these contributions vary with the location of the Sun on the Galactic plane it turns out remarkable that one may predict what will be the Galactic component producing the most tidal effect on the Oort Cloud analogs around new extra solar planetary systems, simply looking at the position of its central star on the plot of Fig. \ref{rotc}.    
\subsection{On planar tide contributions in Hill's approximation}
\begin{figure}[htbp]
      \centering
      \resizebox{\hsize}{!}
      {\includegraphics{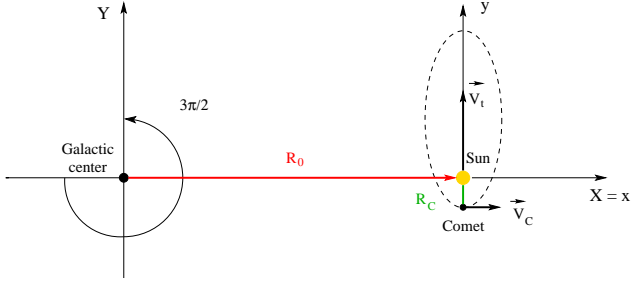}}
      \caption{Picture of the initial location for a comet orbit with galactic longitude $\frac{3\pi}{2}$ in the reference system with the origin on the galactic center (X,Y). The heliocentric system (x,y) (used in Hill's approximation) is also shown (De Biasi's et al.,2015).}
      \label{eliorif}
			\end{figure}

The Hill's approximation (Binney \& Tremaine, 2008) adapts the formalism of the restricted three-body problem (i.e., Solar system, Galaxy, cometary body with mass very small in respect to the previous ones) for the case in which the dimension of satellite system (Solar system) is much smaller than its distance to the center of the host system (Galaxy). In our case the system Sun-comet indeed achieves the maximum size of $1$ pc, while the distances from the Galactic center are about three orders of magnitude larger. In this situation the variation of the gravitational potential along the cometary orbits is very smooth, making possible the application of the \textit{distant-tide approximation} (Binney \& Tremaine, 2008, Chapt.8) . A solar co-rotating system which has the Sun as origin has been considered in this approximation developed for an axisymmetric potential also in 3D-dimensions by De Biasi et al., (2015). The analysis of the comet motion in this \textit{synodic} system is taken into account. The \textit{x-y} plane coincides to the stellar orbital plane, $\mathbf{\hat{e}_x}$ points directly away from the center of the host system and $\mathbf{\hat{e}_y}$ points in direction of the orbital motion of the satellite (Fig.\ref{eliorif}). The reference system rotates with the frequency $\mathbf{\Omega_o}\equiv \Omega_o\mathbf{\hat{e}_z}$.
The most relevant equation in it is the following one:
\begin{equation}
\label{hillmoto}
\ddot{x}(t)= 2\Omega_0\dot{y}(t)+[\Omega_0^2-\Phi''(R_0)]x(t)-\frac{\partial\Phi_s}{\partial x};\\
\end{equation}
giving the acceleration per unit mass of test particle at the point $x(t),y(t)$, along the $x$-axis.
The first term on the right comes from the Coriolis' force, the last one is the contribution of Solar force, the middle one describes the variation of centrifugal force (term $\Omega_o^2$ in square brackets) not completely in balance with the variation of gravitational force due to the host system (Galaxy, term $\Phi''(R_o)$) on the test particle position considered. Assuming a circular motion of the Sun around the Galactic center at the distance $R_o$ and introducing the Oort's constants $A(R), B(R)$, the angular speed of Galactic rotation, measured at $R_o$ is given by:
\begin{equation}
\label{oortc}
\Omega_o=A(R_o)- B(R_o)
\end{equation} 
Then the $x$-component of tide from each Galactic $i$-component ($i=BG,D,DM$, bulge, disk, DM halo) is respectively given by: 
\begin{equation}
 \label{ratid}
 \Omega_{oi}^2-\Phi_i''= \left(-2\Omega_i \Omega_i'R\right)_{R_o}=2\frac{v_{ci}^2}{R_o^2}\left(1-\frac{dln\ v_{ci}}{dln\ R}\right)_{R_o}
 \end{equation}
The main result is that the $x$-component of Galaxy tide turns to be depending on the circular velocity contribution of each Galactic component and on its logarithmic gradient. Conversely the $y$-component turns to be zero due to the balance at Sun position: $\Phi'(R_o)=R_o\Omega^2_o$.\\
That is one of the best results in De Biasi's et al. (2015) work: we know the tidal acceleration per unit length for each unit cometary mass coming from every dynamical component of Galaxy (Eq.\ref{ratid}) before performing integration.

\subsection{Effects of Galactic planar tide}
One of the major effects of the Galactic planar tide on comet's orbit is the perihelion variation which may consist either in the precession of the major axis or in the variation of perihelion distance. The conditions that could amplify or reduce the effects of tidal perturbations turn out to be depending on the three main parameters: the star distance from the Galactic center, the Galactic longitude and the direction of motion of the comet's orbit.

\subsubsection{Testing on comet's orbits}
As exemplification, a test particle has been considered chosing a comet belonging to the outer shell of the Oort Cloud, where the solar gravitational force is lower and the galactic perturbations are then more evident. According to the differential distribution of the inverse semimajor axis for Long Period Comets of Oort's Cloud (Oort, 1950), with a spike at $3.7 \cdot 10^{-5} AU^{-1}$ corresponding to an aphelion of $54000$ AU , the comet considered has initial aphelion of $140000$ AU, initial perihelion of $2000$ AU, inclination on the Galactic plane equal to zero,  Galactic longitude equal to $\frac{3\pi}{2}$  and direct motion direction (Fig.\ref{eliorif}). Two distances are taken into account: $8$ kpc and $4$ kpc from the Galactic center.  


 \begin{figure}[htbp]
 \centering
 \resizebox{\hsize}{!}
{\includegraphics{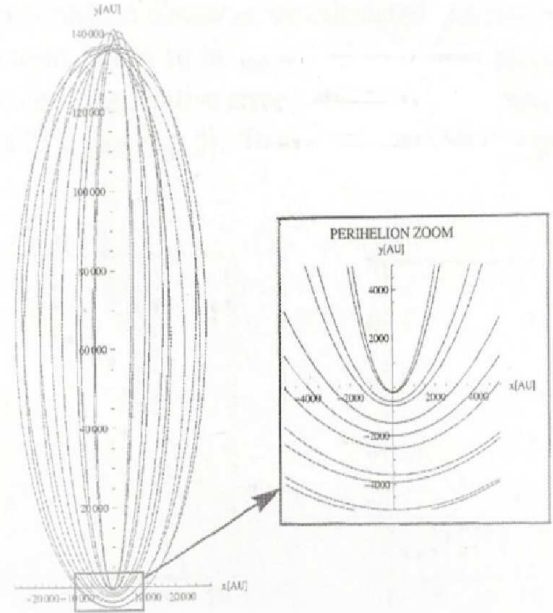}}
\caption{Description of cometary orbits affected by Galactic tides in Hill's approximation when the central star is at $8$ kpc from the center. Perihelion's zooms are also shown in the box (De Biasi, 2010).}  
\label{ZOOM}
\end{figure}
\begin{figure}[htbp]
 \centering
 \resizebox{\hsize}{!}
{\includegraphics{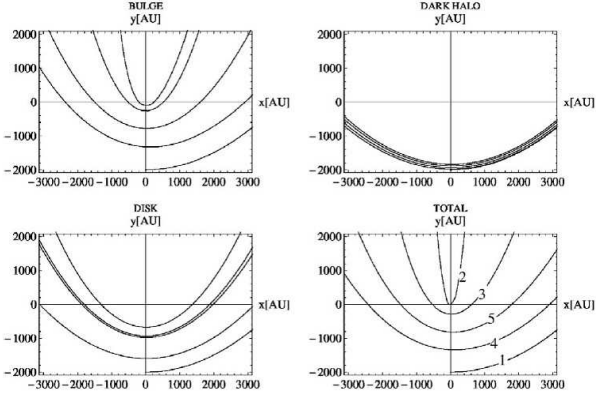}}
\caption{Zoom of the perihelion zone for the comet's orbit considered at $4$ kpc from the Galactic center (Hill's approximation). The most important perihelion's reduction is due here to the bulge. In an integration time of $1$ Gyr, the global effect is to reduce perihelion's distance of $65$\%. The order of time-sequence is marked by numbers in the "Total" case (De Biasi's et al., 2015).}  
\label{EFFBULGE}
\end{figure}

As soon as the Galactic tidal effect is switched on zooms in perihelion zone (Fig.\ref{ZOOM}) are performed distinguishing the contributions from the different dynamical Galactic components. A result in which the effect of bulge is dominant, appears in Fig. (\ref{EFFBULGE}).  

\section{A new constraint on the GHZ}
\begin{figure}[htbp]
 \centering
 \resizebox{\hsize}{!}
{\includegraphics{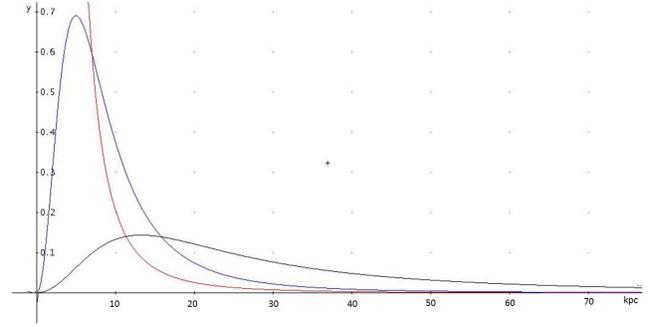}}
\caption{Tidal acceleration from Galaxy components for unit length and unit mass ($x$-component, Eqs.\ref{hillmoto},\ref{ratid}): from bulge (red), disk (blue), DM halo (dark), vs distance from the Galactic center (in kpc), using the contributions of Fig. \ref{rotc}. Bulge dominates until about $7$ kpc, disk in about the range: $7-15$, DM halo over $15$ kpc. An amplification factor of $10^3$ has been used along ordinate axis (De Biasi et al., 2015).}  
\label{TREMAREEPUBBL}
\end{figure}

\begin{figure}[htbp]
 \centering
 \resizebox{\hsize}{!}
{\includegraphics{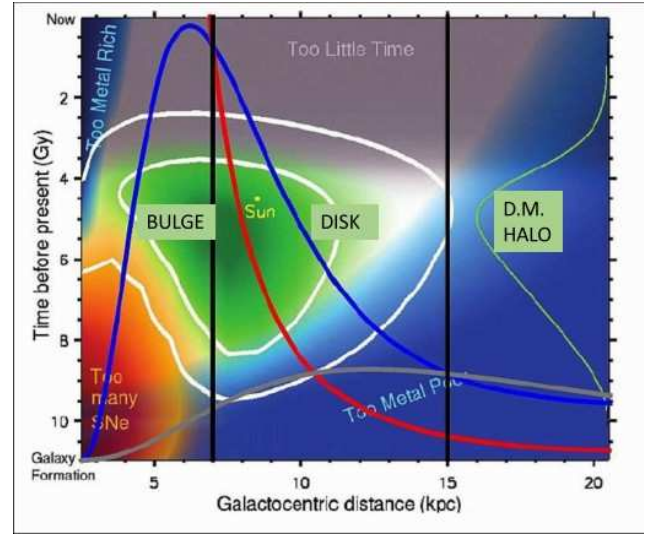}}
\caption{Overlapping of the plot (Fig.\ref{TREMAREEPUBBL}) on the Fig.\ref{IM1} given by Lineweaver et al. (2004). A new relevant factor is added in order to define the GHZ. The domains of different Galaxy components on tide are marked by vertical lines (Fecchio, 2016).}  
\label{MAREEVITA}
\end{figure}

According to the previous result of sect.12.2, we are able to know the contribution to tidal acceleration per unit length and unit mass coming from each Galaxy's component only knowing the contribution of the component considered to the circular velocity (Fig.\ref{rotc}), independently of the central star mass, longitude and direction of motion on comet's orbit. We only need the position of central star on the Galactic plane (Fig.\ref{TREMAREEPUBBL}). 
Then it becomes very interesting to overlap this plot on that of Lineweaver et al., 2004 (Fig.\ref{IM1}) in order to include a new relevant factor on the GHZ discussion (Fig.\ref{MAREEVITA}).
Using Fig.\ref{TREMAREEPUBBL} a quantification may be derived for Fig.\ref{MAREEVITA}.  We immediately may do, e.g. an estimation of the ratio, $X$, of the dominant tide coming from bulge, e.g., if Sun moves at $3$ kpc, in comparison with that due to the dominant disk when Sun is, as now, at $8$ kpc. X turns to be equal to $13.5$. The same ratio becomes $X=1.8$ as soon as the Sun was placed at $6$ kpc, again under the dominant tidal effect from the bulge. Moreover, the relative amount of one Galaxy component over the other, may be given, e.g., at $4$ kpc the tidal bulge effect surpass that of the disk at the same position of a factor of about $50$.\\
In conclusion, if Sun during its probable past migration reached a location on Galactic plane under $7$ kpc, the strong dominating tide from the bulge was probably able to induce cometary injection toward the inner part of the planetary system, producing a cometary flux with possible impacts on it. If Sun conversely remained in the range $7-15$ kpc it would have felt a lower tidal effect from the disk and over $15$ kpc an even weaker effect from DM halo. The considerations are worth also for analog extra solar planetary systems.

\section{On probable solar path through the Galaxy}
\begin{figure}[htbp]
 \centering
 \resizebox{\hsize}{!}
{\includegraphics{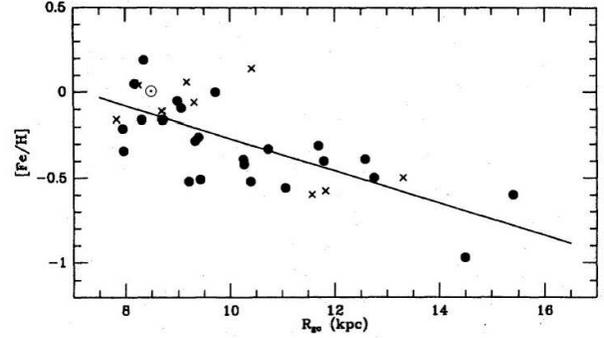}}
\caption{Metal abundance gradient in the Galactic disk for open clusters older than $1$ Gyr, filled circles are from Friel \& Janes (1993), crosses from Lynga (1987). The line corresponds to the linear least squares fit equal to $-0.095 \pm 0.017$ \ dex/kpc (Pagel, 1997, pg.93). Sun's position is also given.}  
\label{PAGEL_FRIEL}
\end{figure}

An important question is: in which location within the Galaxy the Sun formed? The Sun's metallicity actually seems to be over the mean value (of about $20\%$ in $Z$) at the current Sun's position  as derived from the mean metallicity gradient in the Galactic thin disk (Fig.\ref{PAGEL_FRIEL}). This suggests that the Sun  could actually be born at a shorter distance from Galactic center and that it may have undergone migration toward its present location. But how is it possible to reconstruct 
its path only on the basis of metallicity distribution on the disk?\\

\subsection{On the metallicity gradient of Galactic thin disk}

Theoretically speaking, to get
the metallicity gradient of Galaxy disk is not an easy task. The simplest model has to deal with : i) the history of formation of a disk Galaxy inside the cosmological environment, ii) its dynamical evolution, iii) the local star formation prescription, iv) the standard chemical evolution once assumed a given initial mass function, and a model for spectral evolution of stellar populations.\\
If on one side the description of so many physical processes composing the whole picture requires the introduction of many parameters, from the observational point of view there is a very large spread in the data which is perfectly consistent if stellar migration occurred across significant galactocentric distances. Scattering with transient spiral arms (Ro$\tilde{s}$kar et al., 2008 a) and b)) might have caused it.  Paradoxically, the description of Sun's migration meets an intrinsic barrier from observations.\\
There are indeed many good models considering the cosmological, dynamical and chemical evolution (Pagel, 1997; Prantzos and Silk, 1998; Portinari and Chiosi, 1999; Chiappini et al., 2001; Fenner and Gibson, 2003; Naab and Ostriker, 2006; Piovan et al. 2011) which are able to reproduce in a self-consistent way the mean metallicity distribution function (MDF) and the age- metallicity relationship (AMR). However, there is evidence that a large amount of scatter is present in the observed AMR of field stars and open clusters probably due to the superposition of stellar migration.

The overall observational uncertainties, which enormously increase considering the different chemical  species (see, e.g., Table 1 in Portinari \&  Chiosi, 1999), give the impression that to refer to one experimental mean metallicity gradient could be meaningless.\\
In spite of that the different models considered lead to an intrinsic self-consistency, i.e., they are able to point out a mean value for the metallicity gradient on the disk without a so large spread among the models. In order to make a reasonable exercise on Sun's migration we assume that this consistency is really founded representing something intrinsic of disk as soon as we subtract the next occurrence of migration. 

As reference we adopt the paper of Naab and Ostriker (2006) (NaO, hereafter) and compare their results with other relevant contributions to the topic, in particular that of Lineweaver et al. (2004) (LiN, hereafter) referring to the Galaxy evolution model of Fenner \& Gibson (2003).\\
NaO obtain, at the present time, a mean metallicity gradient nearby the Sun, for gas (not too much different from that of stars) equal to:
$$\frac{dlogZ}{dR}=-0.046\ dex\cdot kpc^{-1} $$
The result is about the same given by LiN (interpolation of their Fig.1), leading to a mean value:
	$$\overline{\frac{dlogZ}{dR}}=(-0.05 \pm 0.01) \ dex\cdot kpc^{-1},$$ 
They both predict that the gradient was
significantly steeper in the past at epoch of solar system formation, with a fairly good agreement. The mean value\footnote{It should be noted that also the observed mean logarithmic gradient of oxygen over $H$ for Galactic $HII$ regions, is believed by Pagel (1997, pgs. 93, 227) to be: $ -0.07 / dex \cdot kpc^{-1}$. The same value is also given by Mo et al. (2010, pg.542) to characterize the trend of Milky Way metallicity. In the first determination of GHZ this value has also been used by Gonzalez et al., 2001.} 
between them turns out to be:\\

 $$\overline{\frac{dlogZ}{dR}}=(-0.07 \pm 0.01) \ dex\cdot kpc^{-1}$$

By considering the lower and upper limits of both mean gradients and their accuracy, an average over-metallicity of about $0.1 \ dex$ yields a possible range for the initial Sun's position given by:\\ 

$R_{i\odot}= 5.5\div 6.7$\ kpc\\

By inspecting Fig.\ref{IM1} we notice that the lower value of the initial Sun's position does not represent a problem because it is within the GHZ given by Lineweaver et al. (2004). The only warning comes from tidal trends of Fig.\ref{TREMAREEPUBBL}. It indeed appears in Fig.\ref{MAREEVITA} that under $7$ kpc the tide due to the bulge increases dramatically, causing probably a strong comet bombardment at the beginning of this assumed Solar path.

\subsection{To look for possible migration of Sun-like stars}
Kaib et al. (2011) used Galaxy simulations to approximate the dynamics that Sun-like stars may have experienced during the past $4$ Gyr in the Milky Way starting from the epoch at which the Sun was born. The key ingredient for the processes there described is the presence of transient spiral arms. Solar analog stars at the last timestep of simulations, which starts at $t=10 \ Gyr$, are choosen on the basis of stellar age, position, and kinematics (see, Fig.\ref{TABKAIB}).

\begin{figure}[htbp]
 \centering
 \resizebox{\hsize}{!}
{\includegraphics{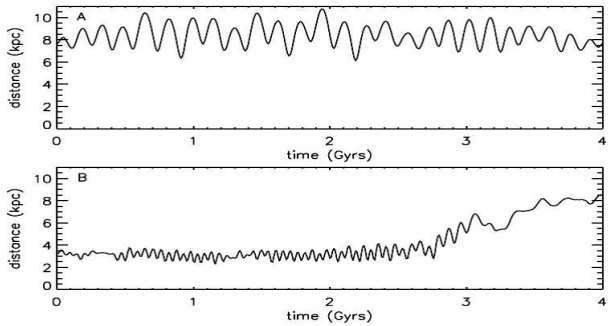}}
\caption{Dynamical history for two solar analogs. One (case A) starts at $8$ kpc at the beginning of simulation and comes back to the same orbital distance at the end of it, oscillating during the last $4$ Gyr between 6 and 10 kpc. The other one (case B) spends most of its first $3$ Gyr orbiting between $2.5-3.5$ kpc from the Galactic center and finally migrates outward to $8$ kpc in the last Gyr of its history (see, text) (Kaib et al., 2011).} 
\label{KAIB}
\end{figure} 

\begin{figure}[htbp]
 \centering
 \resizebox{\hsize}{!}
{\includegraphics{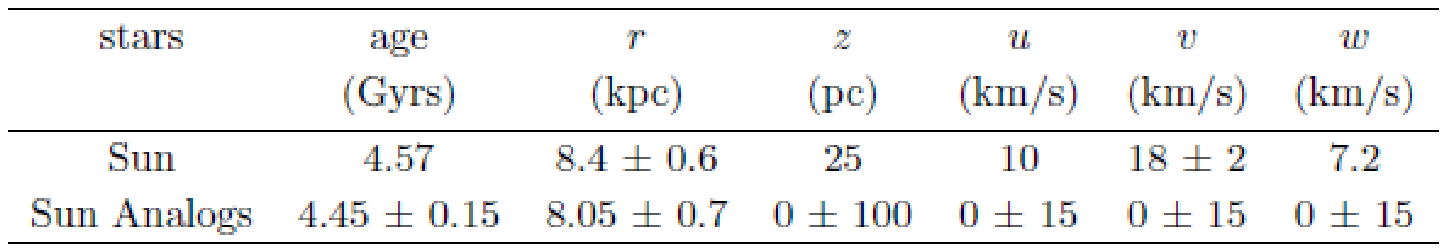}}
\caption{Comparison between the kinematical properties of Sun and those of solar analogs in the Kaib et al. (2011) simulations. Solar distances to the Galactic center and above the midplane so as Solar velocity data are taken from references given in Kaib et al., 2011.}  
\label{TABKAIB}
\end{figure} 

The set of solar analogs dispays a large variety of orbital histories.Two of them are shown in Fig.( \ref{KAIB} A,B).

\subsection{Overlapping possible migration paths on GHZ}
\begin{figure}[htbp]
 \centering
 \resizebox{\hsize}{!}
{\includegraphics{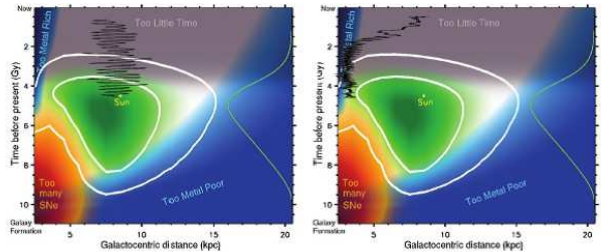}}
\caption{Possible migration paths A and B of Kaib's et al. (2011) simulations are superimposed on the GHZ plot of Lineweaver et al. (2011). Path A, on the left, appears to be compatible also with complex life whereas path B, on the right, does not (see, text). The truncation at the end of both paths is due to the shorter age assumed for Sun (4 Gyr) in the simulations (see, De Biasi, 2014).}  
\label{PATH}
\end{figure}

In the Fig.\ref{PATH} the potential paths A and B obtained by simulations of Kaib et al. (2011) for two Sun's analog stars are superimposed to the GHZ of Lineweaver et al. (2011). The case A (shown on the left) which does not take into account the Sun's initial position due to its over-metallicity, turns out to be compatible with the requests of complex life (and, a fortiori, of not complex). Conversely, case B (on the right) which makes extreme the condition of over-metallicity, appears to be completely in disagreement with the conditions required for developing complex life. Only not complex life is compatible with this path in the last Gyr. However it holds again the warning related to shorter distance than $7$ kpc, seen in sect.14.1, for possible comet's flux during more than the first $3$ Gyr. 

\subsection{Comet's injection and the problem of water}
\begin{figure}[htbp]
 \centering
 \resizebox{\hsize}{!}
{\includegraphics{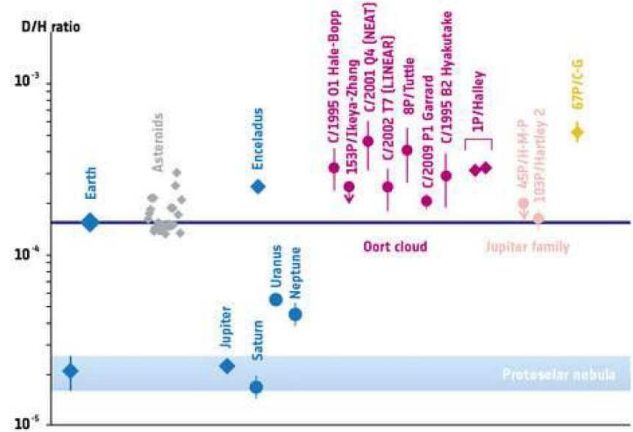}}
\caption{Comparison of values of D/H ratios in different objects in the Solar System, grouped together by color: planets and satellites (blue), chondrites in the asteroid belt (grey), comets from the Oort Cloud (violet), and Jupiter family comets (pink). The comet 67P/Churyumov-Gerasimenko (yellow) has a different D/H ratio from the other comets in the same family. Horizontal line corresponds to the ratio for terrestrial oceans. (Credit: B. Marty/ESA/Altwegg et al., 2014; see, Fecchio, 2016).}

\label{WATER}
\end{figure}

A conclusion about the path B and, in general, about the exclusion of possible locations for planetary systems at distances shorter  than $7$ kpc, due to the stong tidal effect from bulge, cannot be definitively drawn at present time. If a comet's flux toward the inner planets may indeed be produced at these distances, not necessarily it turns out to be dangerous for life but, conversely, it could be necessary for the developing of life. In fact it is still an open problem (Galletta \& Sergi, 2005; Fecchio, 2016): which is the origin of the huge amount of water present on Earth? \\
The high temperature of the protoplanetary disk where the Earth formed (at about $4.5$ Gyr from now) prevented the accretion of large amounts of water. It was probably accumulated later on due to impacts of minor bodies. Asteroids and Jupiter Family Comets appear the most promising candidates for delivering water to the Earth (Fig.\ref{WATER}) even if the amount carried by asteroids might be much less in comparison with that from comets. The discriminating factor is the ratio $D/H$ (deuterium over hydrogen)\footnote{A still open problem is also if the ratio measured in the present ocean water was the same also in the past.} which would rule out comets from Oort's Cloud but also some belonging to the Jupiter Family like the well studied comet 67P/Churyumov-Gerasimenko. The Orbiter Spectrometer for Ion and Neutral Analysis (ROSINA) on board of spacecraft Rosetta (ESA mission 2004-2016) measured for it a D/H ratio about three times greater than that on Earth. Also for the comet Hale-Bopp (1997; period of 2534 yrs) coming from Oort's Cloud and studied by Giotto spacecraft, its ratio turns out to be about two times that of Earth oceans. Conversely the spectral analysis performed by the spatial telescope Herschel (ESA) on comet Hartley 2 (103P/Hartley; period of about $6.5$ yr) of Jupiter Family coming from the Kuiper's Belt, has given a ratio $D/H$  more compatible with that on Earth (Fig.\ref{WATER}). 

In the context of bodies coming from Kuiper's Belt, the presence of a giant planet on an external orbit may be considered either positive or negative. It may act as
a shield against an intense cometary flux due to the presence of an external
perturber exciting cometary eccentricities. In this scenario, like
in the solar system, an external planet (Neptune) may perturb an outer
planetesimal belt (the Edgeworth-Kuiper belt) and inject icy
bodies in the inner regions of the planetary system.
These excited bodies might constantly impact a terrestrial planet
orbiting close to the star possibly in the habitable zone.
However, a giant planet like Jupiter can reduce the probability of
impact of these planetesimals/comets on the inner body by increasing their inclination.
The impact probability is diminished being
inversely proportional to the  $sin (I)$. On the other hand,
if an external perturber(s) capable of stirring the bodies of an
outer planetesimal belt is not present, a giant planet is not
needed to protect the inner terrestrial planets. However, in 
this case the delivery of water from the outer region of a 
protostellar disk becomes problematic and only the migration 
of the planet by tidal interaction with this disk, exchanging angular momentum,  may drive 
a water rich terrestrial planet, formed in the outer
parts of the disk, within the habitable zone. Planets formed 
in situ would be mostly rocky and possibly poor in water.

\section{Cosmological Habitability (COSH)}

We don't refer, obviously, to a special, privileged location of the human being within the Universe, because the Copernican Principle holds\footnote{As consequence of Cosmological Principle it follows indeed that from a physical point of view: {\it We are not priviliged observers of the Universe} (e.g., Peacock, 1999).}. We only suggest that it exists: {\it a collection
of facts (sect.14.1) which connect the factors - constraining the main features of cosmos and
its evolution- with life, by which it is possible to infer how strongly the life 
phenomenon is depending on these factors}. We refer to it as WRAP (Weak Reformulated Anthropic Principle; see, Secco, 2009). 

\subsection{Interplay between the cosmological scenarios 
and the development of life}

First  of all we have to stress the principal factors which constrain
the main features of the Universe.
Briefly we can group them into the three following sectors (see, Dallaporta \& Secco,
1993):

-A) the values of the main constants in Fundamental Physics.

-B) The global properties of the universe and its history (e.g., how it expands, how it builds up the chemical elements, etc).

-C) The space dimensionality.

They then need the link with the essential requests for life already considered in sect.2.\\
Generally speaking, a huge collection of relationships proposed by many authors
(e.g.,Barrow \& Tipler, 1986; Dallaporta \& Secco, 1993; Rees, 2002; Barrow, 2003; 
Gingerich, 2007; Secco, 2009) may prove the 
interplay 
between the present cosmological scenarios and the development of life.
Here we will take into account only few exemplifications
for the sectors A) and B) inviting the reader
to refer at the references cited above. The same holds also for understanding how special is
the spatial dimensionality (C)) equal $3$ of our universe, we will not consider here.

\subsubsection{A) The values of the main constants in Fundamental Physics}
\begin{figure}[htbp]
 \centering
 \resizebox{\hsize}{!}
{\includegraphics{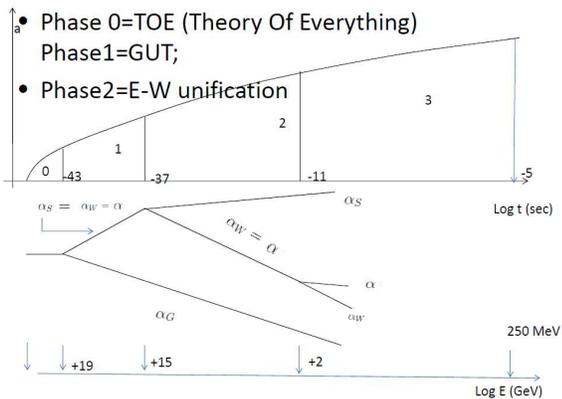}}
\caption{Trends of coupling constants as function of time and energy during cosmological evolution (see, text). On the upper panel the development of expansion parameter is also shown, without inflationary epoch (elaboration from that given in Chapt.6 by Lederman \& Schramm, 1989).}  
\label{COST_ACC}
\end{figure}

We will here refer only to the coupling constant\footnote{Actually they are not constant but function of energy to which they refer (see, Fig. 
\ref{COST_ACC}).} values of the four fundamental forces: $\alpha_G$ (gravitational); $\alpha$ (electromagnetic); $\alpha_W$ (weak); $\alpha_S$ (strong) (Lederman \& Schramm, 1989). The advantage to introduce them lies in the fact that the intensities of corresponding forces are expressed in adimensional terms allowing immediately their comparison. At ordinary energies their values turn out to be: $\alpha_G \simeq 10^{-39}$ , $\alpha\simeq 7.3 \cdot 10^{-3}$ , $\alpha_W\simeq 10^{-5}$, $\alpha_S=15$. In Fig.\ref{COST_ACC} their trends are given for increasing energy going toward singularity.
Even if the reliability of the physical phenomena is decreasing, 
going in this direction, the \textit{electro-weak unification} proved
at about $100~ GeV$ at CERN (Ginevra), allows us to take seriously into account 
the other possible unification as depicted by the grand unified
theory (GUT) at the time of about $\sim 10^{-35}sec$. At this time,
when the energy density of the universe had to be about $10^{15}GeV$,
according to GUT the three forces: the \textit{strong} and the \textit{electro-weak} could
be unified. The spontaneous breaking of the high level symmetry corresponding
to this unified force, occurred owing Universe's expansion, allowed the strong force to separate 
from the \textit{electro-weak} and to assume a coupling constant which now is equal to: $\alpha_S
\simeq 15$. At about $\simeq 10^{-11} sec$ the \textit{electro-weak} symmetry broke too and the
corresponding unified force splits into
the \textit{electromagnetic force} (with a typical coupling constant value of, $\alpha\simeq 1/137$)
and into the \textit{weak} one (with a typical coupling constant of, $\alpha_W\simeq 10^{-5}$).\\ 
These values with which the forces detached differing  one from the other, are crucial in the primordial nucleosynthesis epoch ($\simeq 200 sec$).
Indeed if $\alpha_S$ increases only by $0.3\%$, \textit{dineutron} binds and with, 
$\Delta \alpha_S/\alpha_S$, increasing by $3.4\%$ the \textit{diproton} is bound too. But if, 
$\Delta \alpha_S/\alpha_S$, decreases less than $9\%$, the deuterium nucleus fails 
to be bound (Davies,1972 in Barrow \& Tipler, Chapter 5, 1986, pg. 322). These little changes might have catastrophic consequences for life.
For example, if the deuteron was unbound, the consequences for the nucleosynthesis of
elements necessary for development of Biology, are strong because {\it a key link in 
the chain of nucleosynthesis would be removed} (Barrow \& Tipler, Chapter 5, 1986). Conversely 
if the strong interaction was a little stronger, the \textit{diproton} 
stable bound state would have the consequence that {\it all the hydrogen in the universe would 
have been burnt to $^2He$ (\textit{diproton}) during the early stages  of the Big Bang
and no hydrogen compounds or long-lived stable stars would 
exist today} (Barrow \& Tipler, Chapter 5, 1986, pg. 322). Indeed
the reactions to form $^4He$ would find a channel about $10^{18}$ times faster
in comparison with those without \textit{diproton} formation. The hydrogen reserve would have been
quickly consumed without allowing, e.g., the water formation.\\
Moreover the stability of a nucleus of a mass number $A$ and atomic number $Z$
hinges on a fine link between the strengths of electromagnetic and strong forces as
follows:
$$\frac{Z^2}{A} \le 49 \Big(\frac{\alpha_S}{10^{-1}}\Big )^2\Big(\frac{1/137}{\alpha}\Big)$$
{\it Thus, if the electromagnetic interaction were stronger (increased $\alpha$)
or a stronger 
interaction a little weaker (decreased $\alpha_S$), or both, then
biologically essential nuclei like carbon would not 
exist in Nature} (Barrow \& Tipler, Chapter 5, 1986, pg.326). 

A long collection of other 
exemplifications of this kind are given in the cited book. 

\subsubsection{B) The global properties of universe and its history}
We don't consider here the \textit{fine tuned cosmological expansion} which constraints in a not trivial way the trend of density parameter $\Omega$. That is strictly related to the age of Universe and to the possibility to form structures. Both are of course linked to the necessary conditions for life (e.g., Dallaporta \& Secco, 1993). Neither we will refer about the mistery of the \textit{fine tuning} related to the \textit{cosmological constant $\Lambda$.} Nor we will take into account the interesting answer given by Rees (2002):\textit{"its tuning....may be a fundamental request for our existence".}\\  
Let us remember at least one of the many incredible bottle-necks through which
the Universe's evolutive story passes. It refers to how the Universe build up the chemical elements, precisely how it occurs the production of the so large amount of carbon which life needs.

\subsubsubsection{Carbon and Oxygen nucleosynthesis}
After helium burning, inside stars of first generation, nucleosynthesis transforms three
helium nuclei into one of carbon as follows: $3^4He\rightarrow ^{12}C$. At first two
helium nuclei collide producing the nucleus of $^8Be$. But this nucleus is 
unstable and would decade  in $\simeq 10^{-7} sec$ unless it captures a third helium 
nucleus in order
to change itself into $^{12}C$. But this chain at the reaction temperature of about $10^8K$ would
not produce enough carbon for life unless the last reaction was resonant, that means 
there would exist a level of $^{12}C$ nucleus about equal to the intrinsic energy of the two nuclei
$^8Be+^4He$ plus the mean typical kinetic energy of collision at $10^8K$, so that the reaction rate 
would increase strongly . The resonance level indeed exists and it corresponds to $7.6549 \ MeV$,
as Hoyle predicted since 1954. This resonance channel was soon verified by 
Fowler in the laboratory (e.g., Reeves, 1991, pg.61;
Ortolan \& Secco, 1996). He got the Nobel Prize (1983) also for that. To be noted as Anthropic Principle (weak form, WAP) has again shown its prediction capability as a real physical principle must have.\\
It should be noted that the next reaction of carbon burning by which oxygen
is produced has also to be tuned but in the opposite way. Indeed the following
reaction: $^{12}C+^4He\rightarrow^{16}O+\gamma$ must not to be resonant. If it does,
all the carbon would be transformed into oxygen. Luckily it does not even if
there is a resonant level for the oxygen nucleus but at slightly lower energy, $7.1187 \ MeV$ (Dallaporta \& Secco, 1993). So comparable
quantities of carbon and oxygen are produced to make the $CO$ molecule a
common one. As consequence the formaldehyde $H_2CO$ is simply the association of two of the 
most common molecules 
in the universe ($H_2$ and $CO$). Moreover molecules as:
$$(H_2CO)_n\rightarrow ~sugars~ and~ carbohydrates$$ are easily built up
(Hoyle, 1991).\\
To be noted that the carbon and oxygen energy nuclear levels are 
strictly depending on the values which $\alpha$ and $\alpha_S$ properly have.
If $\alpha$ would vary more than $4\%$ or $\alpha_S$ more than $0.4\%$, the carbon 
or oxygen production will change of a factor in the range $30-1000$ (Barrow, 2003).   

\section{Conclusions}
Starting from the local scale, life leads to connect us with the largest scale, that of Universe. From this analysis a possible scenario arises in which links among the different scales are advanced. Even if possibly partial, a large set of minimum conditions has been identified which must be met for allowing life. The consequence of these conditions is that if we look at life from the probability point of view and then regard it as a complex phenomenon composed, by compatible and independent events, the probability to get it tends drastically to zero. But \textit{we are}! With F. Hoyle we too may then do the following considerations: are we dealing with a  {\it "monstrous sequence of accidents"} ? Conversely "...\textit{my apparently random quirks have become part of a deep-laid
scheme.}
\footnote{{\it "I do not believe that any scientist who examined
the evidence would fail to draw the inference that the laws of nuclear physics have been deliberately
designed with regard to the consequences they produce inside the stars.
If this is so, then my apparently random quirks have become part of a deep-laid
scheme. If not then we are back again at a monstrous sequence of accidents."} (Hoyle, 1959).}"\\
Moreover, regarding some complex phases of the cosmological evolution (e.g. the breakings of symmetries)  it appear particularly surprising the direction they finally undertake (e.g. the corresponding values of coupling constants). Life will be connected to those values but life will be present only many Gyr later. Universe appears to have played a football game but without ball (=life).
Whatever may be what one believes, there is some Beauty in all that, not too much different from what we experience in the current life, when we realize as the possibility to be live at any time is guaranteed by having our blood with many of its parameters together inside very strict ranges. 
Is really finely tuned the architecture within which life, as precious stone, is embedded.      

 \acknowledgments\\
We are grateful to Profs.J.I. Lunine and D. Chernoff of Cornell University for the fruitful collaboration especially during the development of the sections on GHZ. For a general overview we are indebted to Prof. Brad K.Gibson Director of E.A. Milne Centre for Astrophysics at Hull University. Thanks also to Prof. J.F. Kasting who helped us to significantly improve the sections on CHZ. 

\newpage
\section{Appendix}
Nitrogen constitutes about $2.5 \% $ of the human body and about $78 \%$ of the Earth's atmosphere. It plays an essential role in the chemistry of life and it is one of the elemental building-blocks of amino acids. As a consequence, its compounds are fundamental for living organisms.\\
Where does this element come from? In stars of second generation during main sequence phase when the temperature inside is higher than about $15\cdot 10^6 \ K$, by CN cycle. In it the mean reaction times differ from reaction to reaction enormously (e.g., $^{14} N$ has a reaction time  twenty times longer than $^{12} C$). Only when equilibrium is reached, each isotope is formed just as fast by one reaction as it is destroyed by another one. The equilibrium conditions, when the reaction rates per cubic centimeter must be equal for all six steps of the cycle, corresponds to this mass abundance ratios (Schwarschild, 1958):
$$\frac{X(^{14}N+^{15}N)}{X(^{12}C+^{13}C)}= 21\pm 8$$ 
Then the carbon has to be decreased in favor of nitrogen (and oxygen) which becomes abundant within stars.\\
 Another evolutionary phase in which carbon and oxygen too decrease in favor of nitrogen is during the AGB (asymptotic giant branch) for stars of $M\ge 4-5 M_{\odot}$ by the CNO cycle. It involves two groups of reactions: the previous six of CN plus other nine when, $^{15}N+p$, opens  the other low probability (1/2500) channel, going into $^{16}O+ \gamma$, at an higher temperature of about $20\cdot 10^6 K$ (Chiosi, 2012, pg.205; Galletta, 2012). During advanced evolutionary stages, it occurs indeed  that $H_e$-burning in the core is exhausted and the energy star sources are located into two shells, the inner one by $H_e$- burning the outer one by $H$-burning. It follows some phases of three \textit{dredge-up} phenomena, thermal pulses and mass-loss. Then the temperature on the basis of the convective envelope may become so high to start the H-burning by the CNO cycle via a process of HBB (\textit{hot bottom burning}) which allows the star to become the most efficient source of nitrogen (Ferranti, 2015; Chiosi, 2012, pg. 379).

An other essential element for life is the phosphorous. In human body it represents about $1.1 \%$ of body weight. Most of it ($80-85\%$) is in the skeleton while $1\%$ in the brain. It is necessary for nucleic acids and it is involved (with also sulphur) in "high energy" bonds which can supply energy for biochemical reactions (Barrow \& Tipler, 1986, pg.553). It enters indeed methabolic processes within the cells in particular in those transforming chemical into biological energy. Its percentage  by weight in the lithosphere is only of $0.12\%$ (Barrow \& Tippler, 1986, pg.542). That follows from the low abundance of P in the interstellar gas: about 2 atoms of P for each 100 millions of H atoms. The reason of this scarcity is connected with the nuclear synthesis mechanism of phosphorous which occurs by oxygen burning in the core of massive stars ($\ge 12 \ M_{\odot}$) at temperature of about $1.5\cdot 10^9 \ K $ (Chiosi, 2012, pg.215; Galletta, 2012). Massive stars are  indeed less numerous than the low mass ones.\\
Lingam \& Loeb (2017b) have considered the perspective for life on planets with subsurface oceans (e.g., Europa and Encelado). The main idea they take into account is that planets lying ouside the HZ could not be precluded to life. One important aspect considered is that phosphorous must be available in the form of chemical compounds which have to be soluble and active in liquid water in order to supply the necessary energy for life. This would imply a geochemical phosphorous cycle which, e.g., on the Earth,  has been done in the past by volcanic environments (Galletta \& Sergi, 2005). Their conclusion is that this kind of crucial cycle faces challenges on these planets.\\


\end{document}